%% file: main.tex
\documentclass[lettersize,journal]{IEEEtran}
\usepackage[hidelinks]{hyperref}

\usepackage{orcidlink}
\usepackage{amsmath,amsfonts}
\usepackage{algorithmic}
\usepackage{array}
\usepackage[caption=false,font=normalsize,labelfont=sf,textfont=sf]{subfig}
\usepackage{textcomp}
\usepackage{stfloats}
\usepackage{url}
\usepackage{verbatim}
\usepackage{graphicx}
\usepackage{balance}
\def\BibTeX{{\rm B\kern-.05em{\sc i\kern-.025em b}\kern-.08em
    T\kern-.1667em\lower.7ex\hbox{E}\kern-.125emX}}

\usepackage{pifont}%
\newcommand{\cmark}{\ding{51}}%
\newcommand{\xmark}{\ding{55}}%

\usepackage{tikz}
\usetikzlibrary{dsp,positioning}

\usepackage{graphics}

\usepackage{balance}
\usepackage{amsmath}
\usepackage{cite}

\usepackage{acronym}
\newacro{SNR}{signal-to-noise ratio}
\newacro{STFT}{short-time Fourier transform}
\newacro{MMSE}{minimum mean squared error}
\newacro{STSA}{short-time spectral amplitude}
\newacro{log-STSA}{log-STSA}
\newacro{PSD}{power spectrum density}
\newacro{MSE}{mean squared error}
\newacro{MAP}{maximum a posteriori}
\newacro{SI-SDR}{scale-invariant signal-to-distortion ratio}
\newacro{DNS}{Deep Noise Suppression}
\newacro{ESTOI}{extended short-time objective intelligibility}
\newacro{PESQ}{perceptual evaluation of speech quality}
\newacro{DNN}{deep neural network}
\newacro{POLQA}{perceptual objective listening quality  analysis}
\newacro{VAE}{variational autoencoder}
\newacro{MCMC}{Markov Chain Monte Carlo}
\newacro{VI}{variational inference}
\newacro{MCD}{MC dropout}
\newacro{DE}{deep ensembles}
\newacro{AUSE}{area under the sparsification error}
\newacro{RMSE}{root mean squared error}
\newacro{PESQ}{perceptual evaluation of speech quality}
\newacro{AMAP}{approximate MAP}

\usepackage{tabulary}
\usepackage{xcolor}

\DeclareMathOperator*{\argmin}{arg\,min}

\usepackage[subtle]{savetrees}

\usepackage{caption}

\begin{document}
\title{Integrating Uncertainty into Neural Network-based\\ Speech Enhancement}

\author{Huajian Fang$^{*\dag}${\orcidlink{0000-0002-4058-0391}}\thanks{The authors gratefully acknowledge support from the German Research Foundation DFG under the project CML (TRR 169) and ahoi.digital.}, \IEEEmembership{Student Member, IEEE}, Dennis Becker$^{\dag}${\orcidlink{0000-0002-1437-6127}}, \\Stefan Wermter$^{\dag}${\orcidlink{0000-0003-1343-4775}},~\IEEEmembership{Member,~IEEE}, and Timo Gerkmann$^{*}$\,{\orcidlink{0000-0002-8678-4699}},~\IEEEmembership{Senior Member,~IEEE}
\thanks{The authors are with the $^*$Signal Processing (SP) Group, the $^\dag$Knowledge Technology (WTM) Group, Department of Informatics, Universität Hamburg, 22527 Hamburg, Germany (e-mail: \{huajian.fang; dennis.becker-1; stefan.wermter; timo.gerkmann\}@uni-hamburg.de).}}

\maketitle

\begin{abstract}
Supervised masking approaches in the time-frequency domain aim to employ deep neural networks to estimate a multiplicative mask to extract clean speech. This leads to a single estimate for each input without any guarantees or measures of reliability. In this paper, we study the benefits of modeling uncertainty in clean speech estimation. Prediction uncertainty is typically categorized into \emph{aleatoric uncertainty} and \emph{epistemic uncertainty}. The former refers to inherent randomness in data, while the latter describes uncertainty in the model parameters. In this work, we propose a framework to jointly model aleatoric and epistemic uncertainties in neural network-based speech enhancement. The proposed approach captures aleatoric uncertainty by estimating the statistical moments of the speech posterior distribution and explicitly incorporates the uncertainty estimate to further improve clean speech estimation. For epistemic uncertainty, we investigate two Bayesian deep learning approaches: Monte Carlo dropout and Deep ensembles to quantify the uncertainty of the neural network parameters. Our analyses show that the proposed framework promotes capturing practical and reliable uncertainty, while combining different sources of uncertainties yields more reliable predictive uncertainty estimates. Furthermore, we demonstrate the benefits of modeling uncertainty on speech enhancement performance by evaluating the framework on different datasets, exhibiting notable improvement over comparable models that fail to account for uncertainty.

\end{abstract}

\section{Introduction}
\noindent Speech recorded in noisy environments is often corrupted by background noise, which renders it difficult to understand by either humans or machines via automatic speech recognition systems. These problems call for robust speech enhancement algorithms, which extract desired clean speech from noisy mixtures to improve speech quality and intelligibility of recordings. In this paper, we consider single-channel speech enhancement.

Speech enhancement algorithms typically utilize the \ac{STFT} to transfer the recorded signal into the time-frequency domain, where multiplicative filters can be applied to obtain an estimate of clean speech~\cite{timo2012transac, timowienerfiltering2018}. Various Bayesian estimators, e.g., \ac{MAP} and \ac{MMSE} estimators, have been developed based on different statistical distributions about speech and noise, aiming to restore either the spectral coefficients of the \ac{STFT} or the spectral magnitudes~\cite{mmse1984,wolfe2003efficient, Ioannis2006supergaussian, Breithaupt2008pamaetrizedMMSE}. Given the assumption that speech is degraded by uncorrelated additive noise and that both follow complex Gaussian distributions with zero mean, the well-known Wiener filter can be derived. Traditionally, the speech and noise variances estimated by statistical model-based methods~\cite{timo2012transac,martin2001} can be used to construct the \ac{MMSE}-optimal Wiener filter.

Recently, neural networks have been widely used in speech enhancement methods due to their flexibility and effectiveness in nonlinear modeling. Depending on their application, varying degrees of success are reported~\cite{marvin2020spp,mfmvdr2012huang,suhadi2011,deepmmse2020,bandovae,Takuya2021, simonvae,carbajal2021classifier,fang2021variational,carbajal2021disentangle,wang2018supervised,braun2021loss}. Specifically, deep neural networks have been utilized to replace some of the building blocks of conventional speech enhancement methods. For instance, a neural network-based speech presence probability estimator has been proposed in~\cite{marvin2020spp} and combined with a single-channel multi-frame approach~\cite{mfmvdr2012huang}. In \cite{suhadi2011,deepmmse2020}, neural networks are employed to estimate speech and noise power spectrum densities that are required in various Bayesian estimators. Additionally, recent work has leveraged the probabilistic modeling of generative networks for speech enhancement. For example, the \ac{VAE} has been used to estimate the clean speech distribution, which is then combined with a separate noise model to construct a noise reduction Wiener filter~\cite{bandovae, simonvae}. The robustness of this filter can be further improved by injecting noise information~\cite{fang2021variational}, temporal dependencies~\cite{richter2020speech,leglaive2020recurrent,bie2022unsupervised}, and information from other modalities, such as vision~\cite{sadeghi2020audio, carbajal2021disentangle}. Besides, speech enhancement approaches based on perceptual metric-guided adversarial training~\cite{fu2021metricgan+,fu2022metricgan} and diffusion-based generative models~\cite{lu2022conditional,welker2022speech} have also been presented. In contrast, supervised masking approaches~\cite{wang2018supervised} aim to learn a mapping from the noisy input to a masking filter. It allows neural networks to directly estimate a time-frequency filter by training on a large amount of noisy-clean speech pairs using an appropriate cost function~\cite{braun2021loss}. In this work, we focus on supervised masking approaches.

While the time-frequency noise-removing filter aims to remove noise with minimum speech distortions, the algorithm's robustness and reliability are not guaranteed, especially when speech is corrupted by previously unobserved noise. To alleviate this shortcoming, research has been conducted to investigate how to generalize to unseen situations by, e.g., developing more sophisticated network architectures, improved features, or including more training data that covers a wide variety of acoustic scenarios \cite{rehr2021snr,nat2016, largescale2016}. The first is often accompanied by a tremendous increase in model parameters, while the latter is rather time-demanding. Still, improving the generalization ability of neural networks in unseen scenarios is an unsolved problem considering the black-box nature of neural networks. It is thus necessary and beneficial to obtain the associated uncertainty as an indicator of reliability besides the point estimate, especially when the model is processing out-of-distribution samples that are insufficiently represented by training data.

In machine learning, predictive uncertainty is typically decomposed into two categories~\cite{aleaepicML2021,aleaepic2009,kendall2017uncertainties}: \emph{aleatoric} uncertainty and \emph{epistemic} uncertainty. The term aleatoric uncertainty is used to describe the uncertainty of an estimate due to the intrinsic randomness of noisy observations. For speech enhancement, it originates from the stochastic nature of both speech and noise and is reflected in the variance of the clean speech posterior predictive distribution. 
Epistemic uncertainty is of different nature: If the parameters of a neural network are trained, e.g., using different training data, different initialization, or a different number of epochs, different parameters result. Therefore, also the parameters of a neural network used to estimate clean speech are uncertain. This uncertainty of the parameters is called epistemic uncertainty~(also known as \emph{model uncertainty}). For a general introduction to uncertainty modeling, readers are suggested to refer to a review article by H{\"u}llermeier et al.~\cite{aleaepicML2021}. Various uncertainty measures have been employed in the deep regression setting, such as confidence intervals, differential entropy, and variance. Depeweg et al.~\cite{depeweg2018decomposition} propose to measure uncertainty based on the entropy of the predictive distribution, which represents the information level of random variables. Pearce et al.~\cite{pearce2018high} use confidence intervals (which state how certain the estimate is within a certain range) in a distribution-free setting. In this paper, we address uncertainty modeling in a probabilistic way following~\cite{lakshminarayanan2017simple,kendall2017uncertainties,gustafsson2020evaluating} and measure the uncertainty in terms of the \emph{variance}.

\textbf{Aleatoric uncertainty}. 
Due to the stochastic nature of speech and noise, a mapping from noisy speech to clean speech is uncertain as reflected by the posterior predictive distribution of clean speech. We can model this posterior using a specific conditional distribution, such as a Gaussian or a Laplacian~\cite{lakshminarayanan2017simple,kendall2017uncertainties,ilg2018uncertainty}, and employ a neural network to directly estimate the statistical moments of this distribution. While the predicted mean is the \ac{MMSE} estimate of the target \cite{timowienerfiltering2018}, the associated variance can be used to quantify the data inherent uncertainty, i.e., aleatoric uncertainty~\cite{kendall2017uncertainties}.  

Few studies in neural network-based speech enhancement have incorporated the uncertainty of aleatoric nature. Chai et al. propose to use a generalized Gaussian distribution to model the prediction error on a logarithmic scale~\cite{chai2019generalizedgaussian}. In \cite{kinishita2017mdn}, a neural network is used to estimate the parameters of a Gaussian mixture model, which then serves as the basis of an extra statistical model-based speech enhancement approach. This results in only a slight improvement over the baseline optimized with the \ac{MMSE} criterion. Siniscalchi~\cite{Siniscalchi2021distributionalloss} leverages neural networks to learn a histogram distribution to approximate the conditional target speech distribution, which is assumed to be a truncated Gaussian distribution with a fixed variance in each frequency band. However, the fixed variance does not help to capture data-dependent uncertainty.

\textbf{Epistemic uncertainty}. Estimating the statistical moments of the speech posterior predictive distribution allows capturing aleatoric uncertainty, but fails to account for epistemic uncertainty, which corresponds to the uncertainty in neural network parameters~\cite{kendall2017uncertainties,aleaepic2009,aleaepicML2021}. Epistemic uncertainty can be captured using Bayesian inference approaches, which instead of modeling the parameters of a neural network as \emph{deterministic} values, place a distribution over the network parameters and estimates the posterior distribution of the \emph{stochastic} network parameters~\cite{kendall2017uncertainties}. By sampling from the posterior network parameter distribution, multiple sets of neural network parameter realizations can be obtained, thus producing multiple output predictions for each input sample. Uncertainty in predictions due to epistemic uncertainty can be empirically quantified by the variance in these output predictions~\cite{kendall2017uncertainties,aleaepicML2021}. While the true posterior network distribution is intractable~\cite{gal2016dropout}, it can be approximated using 1)~\ac{MCMC} methods~\cite{welling2011bayesian, chen2014stochastic}, which are sampling-based approaches that construct a Markov Chain with the posterior network parameter distribution as its stationary distribution, 2)~variational inference~\cite{blundell2015weight, gal2015bayesian, gal2016dropout}, which approximates the true posterior network parameter distribution with a tractable variational distribution, and 3)~ensemble approaches~\cite{lakshminarayanan2017simple,nixon2020bootstrapped,fort2019deep}, which were proposed from the frequentist perspective but are considered as an approximate Bayesian approach~\cite{gustafsson2020evaluating, wilson2020bayesian}. For instance, Gal et al.~\cite{gal2016dropout} perform variational inference and interpret the dropout regularization technique~\cite{dropout2014} as imposing Bernoulli distributions on the neural network's weights. This method, referred to as Monte Carlo dropout (\emph{MC dropout}), provides a set of target estimates from multiple forward passes by activating dropout at inference. This set of predictions can empirically approximate the outcome distribution for each input sample and allows inference of the variance~(i.e., epistemic uncertainty). In contrast, \emph{Deep ensembles} proposed in~\cite{lakshminarayanan2017simple} can quantify epistemic uncertainty by training multiple neural networks with random weight initialization~\cite{ilg2018uncertainty,gustafsson2020evaluating}.

Recent studies attempt to consider the uncertainty of epistemic nature in, e.g., speech emotion recognition~\cite{mcdropoutser2020, navin2022} and speech recognition~\cite{bbpsr2019, mcdropoutsr202, asrdropout2019}. In~\cite{mcdropoutser2020}, epistemic uncertainty is captured in a speech emotion recognition model for selective prediction, where samples with low confidence (high uncertainty) are rejected. Braun et al.~\cite{bbpsr2019} apply a Gaussian distribution to the weights of an end-to-end speech recognition model to capture uncertainty of neural network parameters, which is then used for parameter pruning. In a recent publication~\cite{mcdropoutsr202}, epistemic uncertainty is employed to improve the robustness of domain adaptation for speech recognition. However, quantifying epistemic uncertainty in neural network-based speech enhancement remains unexplored.

\textbf{Contributions}. Capturing overall predictive uncertainty, which reflects both aleatoric and epistemic uncertainties, is challenging, especially for deep neural networks, but crucial for an understanding of the model's prediction behaviour. In this work, we propose a method that allows capturing aleatoric uncertainty and combining it with epistemic uncertainty approximations to quantify overall predictive uncertainty. In the context of neural network-based speech enhancement, to the best of our knowledge, this is the first work to study different sources of uncertainty in a joint framework and provides for systematic analyses.

We follow the complex Gaussian speech-plus-noise assumption and propose to train a neural network to estimate the Wiener filter and its variance, which quantifies aleatoic uncertainty, based on the \ac{MAP} inference of \emph{complex spectral coefficients}. To regularize the variance estimation, we build an \ac{AMAP} estimator of \emph{spectral magnitudes} using the estimated Wiener filter (mean of the complex clean speech posterior predictive distribution) and uncertainty (variance of the complex clean speech posterior distribution) explicitly. The resulting \ac{AMAP} estimator is in turn used in conjunction with the \ac{MAP} inference of complex spectral coefficients to form a novel hybrid loss function. Rather than discarding uncertainty information at inference, the proposed scheme allows us to explicitly incorporate aleatoric uncertainty approximations into clean speech estimation in a principled way to further correct erroneous speech estimates.

Previous studies on modeling epistemic uncertainty have focused on other tasks than speech enhancement, e.g., \cite{kendall2017bsegnet,ilg2018uncertainty,mcdropoutser2020, navin2022,bbpsr2019, mcdropoutsr202, asrdropout2019}. Yet, questions such as how reliable and accurate the estimates of epistemic uncertainty are in speech enhancement, and how modeling epistemic uncertainty affects enhancement performance, have not been addressed. To this end, we investigate two Bayesian deep learning techniques: MC dropout~\cite{gal2016dropout} and Deep ensembles~\cite{lakshminarayanan2017simple} to capture epistemic uncertainty in clean speech estimation due to their efficiency in approximating Bayesian inference. Although previous works have explored ensemble-based speech enhancement methods~\cite{ensemblejohnatan2013,zhang2017multi}, they did not investigate the effectiveness of ensemble-based methods for uncertainty estimation. 

Moreover, we propose to estimate overall predictive uncertainty reflecting both aleatoric and epistemic uncertainties by combining the proposed hybrid loss function with the ensemble-based method. Finally, we present a comprehensive analysis of uncertainty from different sources and show their impacts on speech enhancement performance over different datasets, which we hope lays the foundation for further use of uncertainties. 

This paper extends our previous conference publication~\cite{fang2022uncertainty}, which studied aleatoric uncertainty. Here, we propose to additionally capture epistemic uncertainty and combine them to quantify overall predictive uncertainty in clean speech estimation. Furthermore, we provide a more detailed analysis with respect to uncertainty estimates from different sources in a joint framework. Section~\ref{sec:signalmodel} describes the signal model. In Section~\ref{sec:aleatoricuncertaintyestimation}, we propose to estimate the uncertainty of aleatoric nature following the complex Gaussian-distributed speech posterior and present how this uncertainty can be incorporated into clean speech estimation. In Section~\ref{sec:bayesianuncertainty}, we show how to capture epistemic uncertainty and quantify overall predictive uncertainty that combines different sources of uncertainty. We introduce the experimental setting in Section~\ref{sec:experiments}, analyze uncertainty estimates in Section~\ref{sec:uncertaintyanalysis}, and present enhancement performance in~Section~\ref{sec:results}. Section~\ref{sec:conclusion} summarizes the findings.

\vspace{-0.1cm}
\section{Signal Model}
\label{sec:signalmodel}
\noindent In the single-channel speech enhancement problem, the noisy mixture consists of clean speech and additive noise. We apply the STFT to obtain the representation in the time-frequency domain as:
\begin{equation}
    X_{ft} = S_{ft} + N_{ft},
  \label{additivemodel}
\end{equation}
where $X_{ft}$, $S_{ft}$, and $N_{ft}$ represent the complex spectral coefficients of mixture, speech, and noise, at the time frame $t\in \{1, 2,\cdots,T\}$ and the frequency bin $f\in\{1,2,\cdots,F\}$. $T$ and $F$ denote the number of time frames and frequency bins respectively. The objective is to recover clean speech in the time-frequency domain by applying a multiplicative filter. To derive such a filter, various assumptions are made according to different signal characteristics. By assuming that the speech and noise coefficients are uncorrelated and follow a circularly symmetric complex Gaussian distribution, 
\begin{equation}
 \label{eq:prior}
S_{ft} \sim \mathcal{N}_\mathbb{C}(0,\,\sigma^{2}_{s,ft}), \hspace{0.3cm}
N_{ft} \sim \mathcal{N}_\mathbb{C}(0,\,\sigma^{2}_{n,ft}) \, ,
 \end{equation}
where $\sigma^{2}_{s,ft}$ and $\sigma^{2}_{n,ft}$ represent the variances of speech and noise respectively, the likelihood $p(X_{ft}|S_{ft})$ follows a complex Gaussian distribution with mean $S_{ft}$ and variance $\sigma_{n, ft}^2$, given by
\begin{equation}
    \label{eq:likelihood}
    p(X_{ft}|S_{ft}) = \frac{1}{\pi \sigma^2_{n,ft}} \exp\left(-\frac{|X_{ft}-S_{ft}|^2}{\sigma^2_{n,ft}}\right) \, .
\end{equation}
With the likelihood in~\eqref{eq:likelihood} and the prior in~\eqref{eq:prior}, we can apply Bayes' theorem to obtain the posterior distribution of clean speech as a complex Gaussian of the form~\cite{timowienerfiltering2018}:
\begin{equation}
    \label{eq:posteriorcoplex}
    p(S_{ft}|X_{ft}) = \frac{1}{\pi \lambda_{ft}} \exp{\left(-\frac{|S_{ft} - W^{\text{WF}}_{ft}X_{ft}|^2}{\lambda_{ft}}\right)} \, ,
\end{equation}

\begin{equation}
 \label{eq:wienerNvariance}
    W^{\text{WF}}_{ft} = \frac{\sigma_{s,ft}^2}{\sigma_{s,ft}^2 + \sigma_{n,ft}^2}, \hspace{0.3cm}
    \lambda_{ft} = \frac{\sigma_{s,ft}^2\sigma_{n,ft}^2}{\sigma_{s,ft}^2 + \sigma_{n,ft}^2}.
 \end{equation}
$W^{\text{WF}}_{ft}$ is recognized as the \emph{Wiener filter} and $\lambda_{ft}$ is the variance of the posterior distribution. Under this assumption, the \ac{MMSE} estimator, which corresponds to the expectation of the posterior distribution, leads to the Wiener filter applied as: 
\begin{equation}
    \widetilde{S}_{ft} = W^{\text{WF}}_{ft} \cdot X_{ft}.
  \label{eq:winerfiltering}
\end{equation}
Due to the symmetry of the complex Gaussian distribution, the \ac{MAP} estimator of complex speech coefficients is identical to the \ac{MMSE} estimator.

\begin{figure*}[htbp]
  \centering
  \captionsetup{justification=centering}
  \input{block_diagram.tikz}
\caption{Block diagram of the proposed neural network-based aleatoric uncertainty estimation.}
\vspace{-0.3cm}
\label{fig:uncertainty_diagram}
\end{figure*}
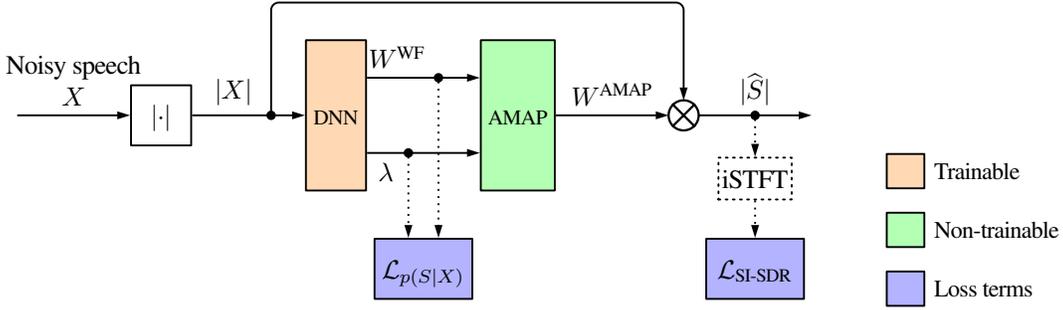 

\section{Aleatoric Uncertainty Estimation}
\label{sec:aleatoricuncertaintyestimation}
\noindent Although speech enhancement is typically formulated as a problem with a single output, the dependency between input and output can be modeled stochastically by means of a speech posterior predictive distribution $p(S_{ft}|X_{ft})$, i.e., a variance $\lambda_{ft}$ is associated with the clean speech estimate and can be interpreted as a measure of uncertainty of the Wiener estimate~\cite{timowienerfiltering2018}. This uncertainty accounts for random effects in data and is referred to as \emph{aleatoric uncertainty}~\cite{kendall2017uncertainties,lakshminarayanan2017simple}. When properly captured, aleatoric uncertainty can reflect the expected estimation error in the absence of ground truth.

\subsection{Deep Aleatoric Uncertainty Estimation}
\noindent In contrast to traditional signal processing techniques~\cite{timo2012transac, timowienerfiltering2018, fevotte2011algorithms}, where the Wiener filter is constructed by separately estimating the variances of speech and noise from the noisy mixture $X_{ft}$, neural network-based supervised masking methods allow direct estimation of multiplicative filters. Besides the Wiener filter $W_{ft}^{\text{WF}}$, one can further estimate the data-dependent aleatoric uncertainty $\lambda_{ft}$ if the neural network is optimized using the speech posterior predictive distribution~(\ref{eq:posteriorcoplex}), i.e., by minimizing the negative logarithm of the posterior distribution of clean speech $p(S_{ft}|X_{ft})$ (the logarithm does not affect the optimization problem due to monotonicity) and averaging over time-frequency bins:
\begin{equation}
    \begin{split}
    &\widetilde{W}^{\text{WF}}_{ft}, \widetilde{\lambda}_{ft} = \\
    &\argmin_{W^{\text{WF}}_{ft},\lambda_{ft}} \underbrace{\frac{1}{FT}\sum_{f,t} \log(\lambda_{ft})  + \frac{|S_{ft} - W^{\text{WF}}_{ft} X_{ft}|^2}{\lambda_{ft}}}_{\mathcal{L}_{p(S|X)}} \, ,   
    \end{split}
    \label{eq:logposterior}
\end{equation}
where $\widetilde{W}^{\text{WF}}_{ft}$, $\widetilde{\lambda}_{ft}$ denote estimates of the Wiener filter and associated aleatoric uncertainty~\cite{kendall2017uncertainties,lakshminarayanan2017simple}.

In contrast, if we assume a constant uncertainty for all time-frequency bins, i.e., $\lambda_{ft} = \lambda^\ast$, and refrain from explicitly optimizing for $\lambda^\ast$, $\mathcal{L}_{p(S|X)}$ degenerates into the well-known \ac{MSE} loss
\begin{equation}
\mathcal{L}_{\text{MSE}} = \frac{1}{FT}\sum_{f,t}|S_{ft}-W^{\text{WF}}_{ft}X_{ft}|^2 \, ,
\label{eq:mse}
\end{equation}
which is widely used in neural network-based regression tasks including speech enhancement~\cite{braun2021loss}. However, neural networks trained to perform point estimation do not necessarily output reliable estimates for clean speech when processing out-of-distribution samples that are underrepresented by the training data~\cite{rehr2021snr}. In this work, we discard the assumption of constant uncertainty; instead, we propose to treat uncertainty estimation as an additional task by training a neural network with the negative log speech posterior $\mathcal{L}_{p(S|X)}$. Consequently, this method not only allows us to obtain a noise-removing mask, but also empowers the model to capture the uncertainty of aleatoric nature associated with predictions.

Modeling aleatoric uncertainty by minimizing the logarithm of the posterior predictive distribution results in an improvement over baselines that fail to account for uncertainty in computer vision tasks~\cite{kendall2017uncertainties}. However, directly using $\mathcal{L}_{p(S|X)}$ as the loss function is prone to overfitting~\cite{fang2022uncertainty} and may result in reduced estimation performance of the Wiener filter. A recent publication~\cite{pitfallnll2022} also reveals that directly minimizing the logarithm of the conditional probability hinders the training of mean estimation, which leads to premature convergence. To tackle this problem, we propose an additional regularization of the loss function by incorporating the estimated uncertainty into clean speech estimation as described next.

\vspace{-0.2cm}
\subsection{Joint Enhancement and Uncertainty Estimation}
\label{sec:proposedscheme}

\noindent Estimating uncertainty $\lambda_{ft}$ associated with the Wiener filter is challenging since ground truth of uncertainty is not readily available. Instead, uncertainty estimation is an unsupervised task with an unspecified search space, which can potentially lead to unstable training~\cite{varvariance2020, reliableNLL2019}. In this work, we propose to incorporate a subsequent speech enhancement task that explicitly uses both the Wiener filter and its uncertainty $\lambda_{ft}$ during the training procedure. The speech enhancement task provides additional coupling between the outputs (Wiener filter and uncertainty). In this manner, the neural network is guided to estimate the uncertainty values that are relevant to the speech enhancement task, as well as to enhance the estimation of the Wiener filter.

Considering complex coefficients with a symmetric posterior~\eqref{eq:posteriorcoplex}, the MAP and MMSE estimators both lead directly to the Wiener filter $W^{\text{WF}}_{ft}$ and do not require an uncertainty estimate. However, this situation changes if we consider spectral magnitude estimation. The magnitude posterior $p(|S_{ft}|\:|X_{ft})$, derived by integrating the phase out of~\eqref{eq:posteriorcoplex}, follows a Rician distribution~\cite{wolfe2003efficient}

\begingroup
\small
\begin{equation}
\begin{split}
    &p(|S_{ft}|\:|X_{ft}) =\\ &\frac{2|S_{ft}|}{\lambda_{ft}} \exp\left(-\frac{|S_{ft}|^2+(W_{ft}^{\text{WF}})^2|X_{ft}|^2}{\lambda_{ft}}\right)\mathit{I_0}\left(\frac{2|X_{ft}|\,|S_{ft}|W^{\text{WF}}_{ft}}{\lambda_{ft}}\right)\,,
\end{split}
\label{eq:rician_speech}
\end{equation}
\endgroup
where $\mathit{I_0}(\cdot)$ is the modified zeroth-order Bessel function of the first kind.

\begin{figure}[b]
\vspace{-0.35cm}
  \centering
  \includegraphics[width=0.65\linewidth]{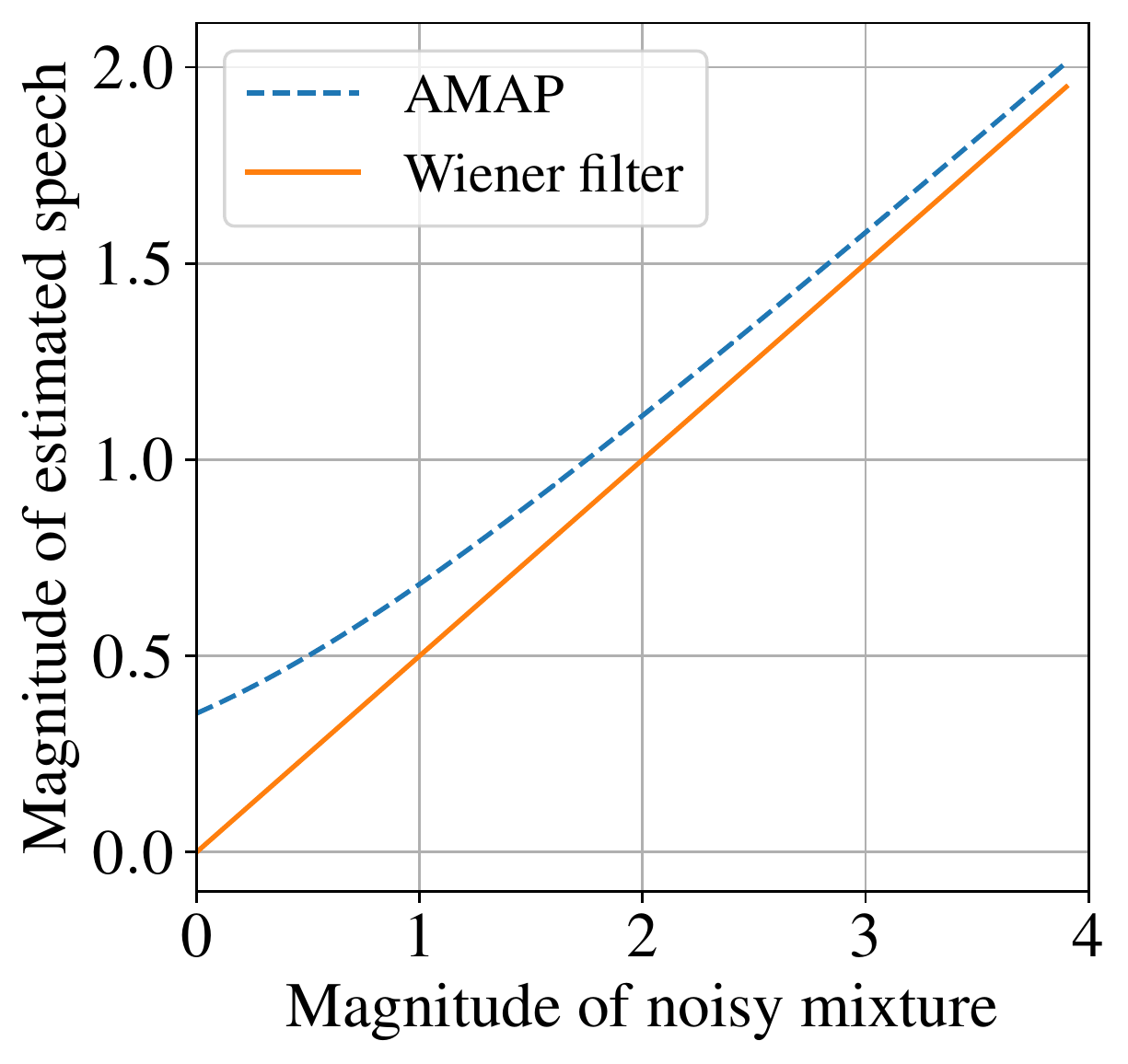}
\vspace{-0.05cm}
\caption{Input-output characteristics of the AMAP estimator $W^{\text{AMAP}}_{ft}$ and Wiener filter $W^{\text{WF}}_{ft}$~(setting $\sigma_{s,ft}^2=\sigma_{n,ft}^2=1$ in this example).}
\label{fig:AMAPWF}
\vspace{-0.35cm}
\end{figure} 

In order to compute the MAP estimate for the spectral magnitude, the mode of the Rician distribution has to be estimated, which is difficult to do analytically. However, it can be approximated by substituting a Bessel function approximation following~\cite{macaulayMalpass1980} into~\eqref{eq:rician_speech} and maximizing with respect to the spectral magnitude, yielding a simple closed-form expression~\cite{timowienerfiltering2018, wolfe2003efficient}:
\begin{equation}
    \label{eq:approximated_map}
    \begin{split}
    |\widehat{S}_{ft}| &\approx W^{\text{AMAP}}_{ft}|X_{ft}|\\
    &= \left(\frac{1}{2}W^{\text{WF}}_{ft} + \sqrt{\left(\frac{1}{2}W^{\text{WF}}_{ft}\right)^2 + \frac{\lambda_{ft}}{4|X_{ft}|^2}}\right) |X_{ft}| \, ,        
    \end{split}
\end{equation}
where $|\widehat{S}_{ft}|$ is an estimate of the clean spectral magnitude $|S_{ft}|$ using the \ac{AMAP} estimator of spectral magnitudes $W^{\text{AMAP}}_{ft}$. It can be noticed that the estimator $W^{\text{AMAP}}_{ft}$ utilizes both the Wiener filter $W^{\text{WF}}_{ft}$ and the associated uncertainty $\lambda_{ft}$. Fig.~\ref{fig:AMAPWF} illustrates the input-output estimation characteristics of the AMAP estimator and Wiener filter~\cite{timowienerfiltering2018}. We can see that $W^{\text{AMAP}}_{ft}$ is \emph{nonlinear} with respect to the noisy input and tends to cause less target attenuation than the Wiener filter especially for low inputs. This indicates that incorporating the associated uncertainty $\lambda_{ft}$ may increase the robustness of the estimator by potentially preserving more speech at the slight cost of noise removal. 

After combining the estimated magnitude $|\widehat{S}_{ft}|$ with the noisy phase, we can apply the inverse~\ac{STFT} to obtain an estimate of the time-domain speech signal, denoted as $\widehat{s}$. Afterwards, the estimated time-domain signal is used to compute the negative \ac{SI-SDR} metric \cite{le2019sdr}: 
\begin{equation}
\vspace{-0.05cm}
\label{eq:sisdr}
    \mathcal{L}_{\text{SI-SDR}} = -10\log_{10}\left(\frac{||\alpha s||^2}{||\alpha s - \widehat{s}||^2}\right)\, , \quad \alpha = \frac{\widehat{s}^{T}s}{||s||^2}\, ,
\end{equation}
which is leveraged as an additional term in the loss function that forces the speech estimate (computed with $W^{\text{AMAP}}_{ft}$) to be similar to the time-domain clean speech target $s$. While a spectrum loss like~(\ref{eq:mse}) is a straightforward solution to regularize the uncertainty estimation, the time-domain loss is expected to be more effective since it is directly related to the raw waveform, implicitly taking phase information into account and thus promoting speech reconstruction for better perceptual performance~\cite{heitkaemper2020demystifying}.

Eventually, we propose to combine the \ac{SI-SDR} loss $\mathcal{L}_{\text{SI-SDR}}$ with the negative log-posterior $\mathcal{L}_{p(S|X)}$ given in~\eqref{eq:logposterior}, and train the neural network using a hybrid loss function
\begin{equation}
    \mathcal{L} = \beta \mathcal{L}_{p(S|X)} + (1-\beta) \mathcal{L}_{\text{SI-SDR}}\, ,
    \label{eq:proposedloss}
\end{equation}
with the weighting factor $\beta \in [0,1]$. By explicitly using the estimated uncertainty for the speech enhancement task, the hybrid loss guides both mean and variance estimation to improve speech enhancement performance. Fig.~\ref{fig:uncertainty_diagram} depicts a block diagram of this approach.

\section{Bayesian uncertainty estimation}
\label{sec:bayesianuncertainty}

\noindent While neural networks performing point estimation have demonstrated effectiveness in speech enhancement, it is not guaranteed that neural networks can generalize well to unfamiliar acoustic situations. Therefore, to quantify the overall predictive confidence regarding the estimated clean speech, it is necessary to also assess the uncertainty of the neural network parameters~(i.e., \emph{epistemic uncertainty}). Note that a single neural network optimized using the proposed hybrid loss~(\ref{eq:proposedloss}) allows capturing aleatoric uncertainty but is unaware of epistemic uncertainty. To solve this, we can utilize Bayesian deep
learning approaches, assuming that the weights of a neural network follow some probability distribution rather than deterministic values. Furthermore, when combined with the loss~\eqref{eq:proposedloss}, an ensemble of networks can provide both aleatoric uncertainty and epistemic uncertainty estimates.

\subsection{Epistemic Uncertainty Estimation}
\noindent Bayesian deep learning provides a set of principled methods to capture epistemic uncertainty~\cite{welling2011bayesian, chen2014stochastic,gal2015bayesian, gal2016dropout,lakshminarayanan2017simple, fort2019deep}. Early work on \ac{MCMC} methods~\cite{welling2011bayesian,chen2014stochastic} constructs a Markov chain with the posterior network parameter distribution as its stationary distribution and generates multiple network parameter realizations by sampling from this distribution. However, \ac{MCMC} methods are computationally inefficient and do not scale well to neural networks with a large number of parameters~\cite{gustafsson2020evaluating,ilg2018uncertainty}. Recent work based on variational inference allows approximating the true posterior network parameter distribution with a tractable distribution~\cite{blundell2015weight, gal2015bayesian}, while at the same time ensemble-based methods are proposed as simple and scalable frequentist alternatives to model uncertainty~\cite{lakshminarayanan2017simple,nixon2020bootstrapped,fort2019deep}. Among the existing Bayesian deep learning methods, MC dropout and Deep ensembles have shown their scalability in large neural network-based problems, such as semantic segmentation~\cite{kendall2017uncertainties} and depth estimation~\cite{gustafsson2020evaluating}. Here, we investigate their effectiveness for uncertainty estimation in speech enhancement.

We define a neural network as a function parameterized by $\theta$ and a training dataset that contains noisy-clean speech pairs $\mathcal{D}=\{(S_{11}, X_{11}),\ldots,(S_{FT}, X_{FT})\}$. Hereafter we omit the indices $ft$, since all time-frequency bins are treated independently in~(\ref{eq:posteriorcoplex}). Since the posterior network parameter distribution $p(\theta|\mathcal{D})$ is computationally intractable in a high dimensional space, variational inference approximates the true posterior network parameter distribution by a pre-specified variational distribution $q(\theta)$ and the speech posterior predictive distribution at inference time is obtained by marginalizing out $q(\theta)$ as:

\begin{equation}
\begin{split}
   p(S|X, \mathcal{D})
  & = \int p(S|X, \theta)p(\theta|\mathcal{D})d\theta \\
  & \approx \frac{1}{M}\sum_{m=1}^{M}p(S|X, \theta_{m}),\quad \theta_m\sim q(\theta),
\end{split}
\label{eq:MCsamples}
\end{equation}
where $\theta_m$ represents $m$-th sampling from $q(\theta)$~\cite{Neal1995BayesianNN}. MC dropout approximates the posterior network parameter distribution using the Bernoulli distribution and samples neural network weights by activating dropout at inference time. Gal et al. provide further details on the derivations in~\cite{gal2016dropout}. This allows obtaining $M$ target speech
estimates from multiple stochastic forward passes for each input. In contrast, Deep ensembles repeatedly train the same model $M$ times with random initialization and random data shuffling~\cite{lakshminarayanan2017simple}, generating $M$ neural networks with deterministic network parameter estimates $\{\theta_m\}_{m=1}^{m=M}$. Since $\theta_m$ can be viewed as independent samples from a certain approximate distribution $q(\theta)$, Deep ensembles can be considered equivalent to approximate Bayesian inference~\cite{gustafsson2020evaluating}. Therefore, the predictive distribution is obtained similarly to~(\ref{eq:MCsamples}). Furthermore, neural networks usually contain a large number of parameters, which makes them multi-modal in the parameter space. Different initialization starting points in Deep ensembles allow the neural network to converge to different local optima, thus potentially capturing multiple modes of $p(\theta|\mathcal{D})$~\cite{fort2019deep, gustafsson2020evaluating}.

Epistemic uncertainty can be approximated by building an ensemble of neural networks using either MC dropout or Deep ensembles, where each network is trained to estimate the Wiener filter only with the loss function~$\mathcal{L}_{\text{MSE}}$~(\ref{eq:mse}). With the results of $M$ forward passes, we can approximate the mean and variance of the distribution $p(S|X)$ by the empirical mean and variance of the prediction set~\cite{gal2016dropout, ilg2018uncertainty}:

\begin{equation}
    \widetilde{S} = \frac{1}{M}\sum_{m=1}^{M} \widetilde{S}_{\theta_m},\quad
 \widetilde{\Sigma} = \frac{1}{M}\sum_{m=1}^{M} |\widetilde{S}_{\theta_m} - \widetilde{S}|^2,
   \label{mean_var_epi}
\end{equation}
where $\widetilde{S}_{\theta_m}$ denotes clean speech estimated using the neural network with parameters $\theta_m$. $\widetilde{S}$ represents the average clean speech estimate and $\widetilde{\Sigma}$ quantifies the epistemic uncertainty.

\subsection{Overall Predictive Uncertainty}
\noindent In the case of optimizing the network using~\eqref{eq:proposedloss}, besides the Wiener estimate $\widetilde{S}_{\theta_m}$, each neural network with weights $\theta_m$ can produce the associated variance $\widetilde{\lambda}_{\theta_m}$. The overall predictive uncertainty, which reflects both aleatoric and epistemic uncertainties, can be computed using the law of total variance~\cite{kendall2017uncertainties,ilg2018uncertainty}:
\begin{equation}
    \widetilde{S} = \frac{1}{M}\sum_{m=1}^{M} \widetilde{S}_{\theta_m}, \quad 
 \widehat{\Sigma} = \frac{1}{M}\sum_{m=1}^{M} \left(|\widetilde{S}_{\theta_m} - \widetilde{S}|^2 + \widetilde{\lambda}_{\theta_m} \right),
   \label{mean_wf_var_pre}
\end{equation}
where $\widetilde{S}$ denotes the average Wiener estimate, and $\widehat{\Sigma}$ quantifies the overall predictive uncertainty.

For each neural network with weights $\theta_m$, we can further generate the \ac{AMAP} clean speech estimate $\widehat{S}_{\theta_m}$ by explicitly incorporating the associated uncertainty $\widetilde{\lambda}_{\theta_m}$ as in~\eqref{eq:approximated_map}. Therefore, given an ensemble of networks, besides the average Wiener estimate $\widetilde{S}$, the average AMAP estimate can be obtained by:
\begin{equation}
    \widehat{S} = \frac{1}{M}\sum_{m=1}^{M} \widehat{S}_{\theta_m}.
   \label{mean_amap}
\end{equation}

\section{Experimental setting}
\label{sec:experiments}

\subsection{Datasets}
\noindent For training and validation, we use a subset of the \ac{DNS} Challenge's training set~\cite{reddy2020interspeech}, which contains synthetic audio samples of 100 hours with \acp{SNR} uniformly distributed between \mbox{-5} dB and 20 dB. The dataset is randomly split into 80 and 20 hours for training and validation respectively. The model is evaluated on two different unseen datasets. The first is the reverb-free synthetic test set released by DNS Challenge. This evaluation dataset is disjoint from the training and validation datasets and is created by adding noise signals sampled from 12 categories~\cite{reddy2020interspeech} to speech signals from~\cite{pirker2011pitch} at SNRs distributed between 0~dB and 25~dB~\cite{reddy2020interspeech}. The second unseen evaluation dataset is created using clean speech from the evaluation subset of WSJ0~(\texttt{si\_et\_05})~\cite{garofolo1993csr} and four types of noise from CHiME3~(\texttt{cafe}, \texttt{street}, \texttt{pedestrian}, and \texttt{bus})~\cite{chime3dataset}. The \acp{SNR} are randomly selected from\,\{\mbox{-10~dB}, -5~dB, 0~dB, 5~dB, 10~dB\}.

\subsection{Architecture and Hyperparameters}
\noindent To ensure a fair comparison, all experiments are performed based on the same U-Net neural network architecture~\cite{Jansson2017SingingVS, tan2018convolutional}. The U-Net structure with skip connections between the encoder and the decoder is comprised of several blocks, each of which consists of: 2D convolution layer + instance normalization~\cite{ulyanov2017improved} + Leaky ReLU with slope 0.2. The encoder contains 6 blocks that increase the feature channel from $1$ to $512$ progressively ($1-16-32-64-128-256-512$), while the decoder reduces it back to 16~($512-256-128-64-32-16-16$), followed by a $1\times1$ convolution layer that outputs a mask of the same shape as the input. For all blocks, the kernel size is set to $(5, 5)$ with stride $(1, 2)$ and padding $(2, 2)$, processing a 2-D input with a dimension of $(T, F)$. For the model estimating aleatoric uncertainty, the output layer is split into two heads that predict both the Wiener filter and associated uncertainty\footnote{Code for the model is available at: \url{https://github.com/sp-uhh/uncertainty-SE}.}. We applied the sigmoid activation function to the estimated Wiener filter, while using the \emph{log-exp} technique to constrain the uncertainty output to be greater than 0, i.e., the network outputs the logarithm of the variance, which is then recovered by the exponential term in the loss function. The batch size is 64; the learning rate is 0.001; the weight decay parameter is set to 0.0005. All neural networks are trained with the Adam optimizer~\cite{adamkinma}. The training process is stopped if the validation loss fails to decrease for 10 consecutive epochs and the learning rate is halved when the validation loss does not decrease for 3 epochs.  

The noisy-clean speech pairs have a sampling rate of 16~kHz, and the \ac{STFT} is computed using a 32~ms Hann window with 50\% overlap.

\subsection{Methods}
\noindent The algorithms considered in this work include:
\begin{enumerate}
    \item \emph{Baseline WF}: The U-Net architecture was trained on noisy-clean speech pairs using loss function~\eqref{eq:mse}. This serves as a baseline, assuming a constant variance for all time-frequency bins and estimating the Wiener filter for each input only.  

    \item \emph{Baseline SI-SDR}: Following the same constant variance assumption as \emph{Baseline WF}, the U-Net network was trained to output a multiplicative filter and optimized using the time-domain loss function~\eqref{eq:sisdr}. This serves as another baseline that fails to account for uncertainty. 
    
    \item \emph{Aleatoric-WF} \& \emph{Aleatoric-AMAP}: The hybrid loss function~(\ref{eq:proposedloss}) allows us to generate two possible clean estimates for each input, i.e., by using the estimated Wiener filter~(\ref{eq:winerfiltering}) or by applying the AMAP estimator~(\ref{eq:approximated_map}) that incorporates both the Wiener filter and its associated uncertainty. They are denoted as \emph{Aleatoric-WF} and \emph{Aleatoric-AMAP} respectively. We observe experimentally that the performance of Aleatoric-AMAP only fluctuates slightly with different $\beta$ values, while the performance of Aleatoric-WF decreases when the value of $\beta$ is large. The weighting factor $\beta$ was empirically chosen to be 0.001 to achieve a good trade-off between the performance of Aletoric-WF and Aleaotirc-AMAP.
    
    \item \emph{MC dropout}: Inserting dropout after each convolution layer regularizes too strongly and impacts the model performance~\cite{kendall2017bsegnet}, which was confirmed in our preliminary experiments. We thus studied several variants of the U-Net by inserting the dropout layer at different positions of the architecture, and selected the variant with three dropout layers~(drop probability of 0.5~\cite{kendall2017bsegnet,gustafsson2020evaluating,dropout2014}) inserted after the three deepest blocks of the encoder. The same cost function as \emph{Baseline WF} is used. This method captures epistemic uncertainty by activating the dropout layers at inference.
    \item \emph{Deep ensembles}: The same setup as \emph{Baseline WF} was trained $M$ times with random initialization. This allows the model to capture epistemic uncertainty.

    \item \emph{DE-Aleatoric-WF} \& \emph{DE-Aleatoric-AMAP}: The same setup as \emph{Aleatoric-WF/AMAP} was trained $M$ times with random intialiation. This allows capturing aleatoric and epistemic uncertainties simultaneously. We average over the estimates according to~\eqref{mean_wf_var_pre} and~\eqref{mean_amap} to obtain two clean speech estimates: \emph{DE-Aleatoric-WF} and \emph{DE-Aleatoric-AMAP} respectively.
    
\end{enumerate}

\begin{figure}[t]
\begin{minipage}[b]{0.25\textwidth}
  \centering
    \centerline{\footnotesize {\hspace{0.8cm}} (a) Noisy}
  \centerline{\includegraphics[width=0.95\linewidth, height=3.6cm]{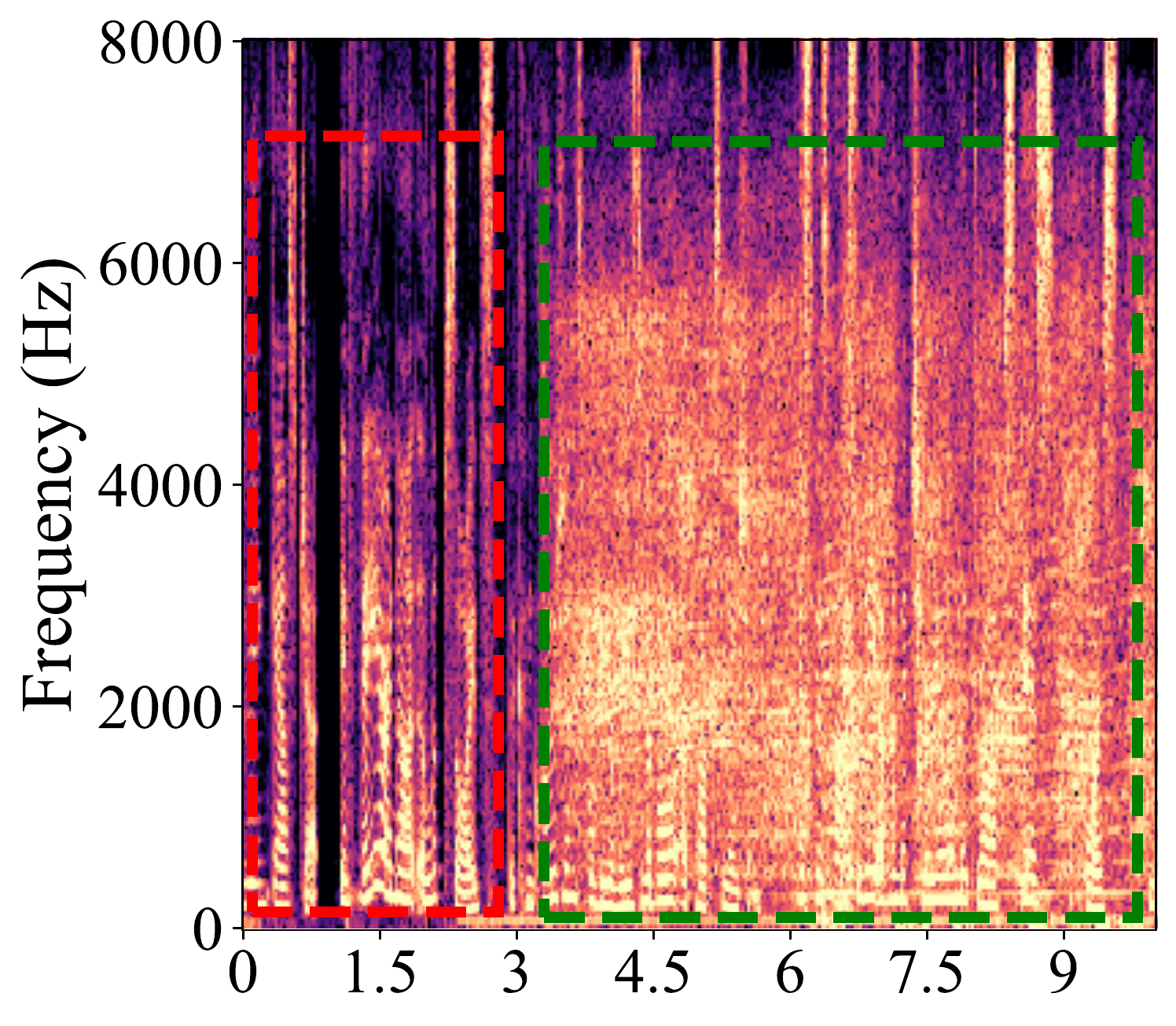}}
\end{minipage}%
\begin{minipage}[b]{.303\textwidth}
  \label{fig:aleatoricerror}
  \centering
    \centerline{\footnotesize (b) Clean {\hspace{0.5cm}}}
  \centerline{{\hspace{-0.15cm}}\includegraphics[width=0.86\linewidth, height=3.6cm]{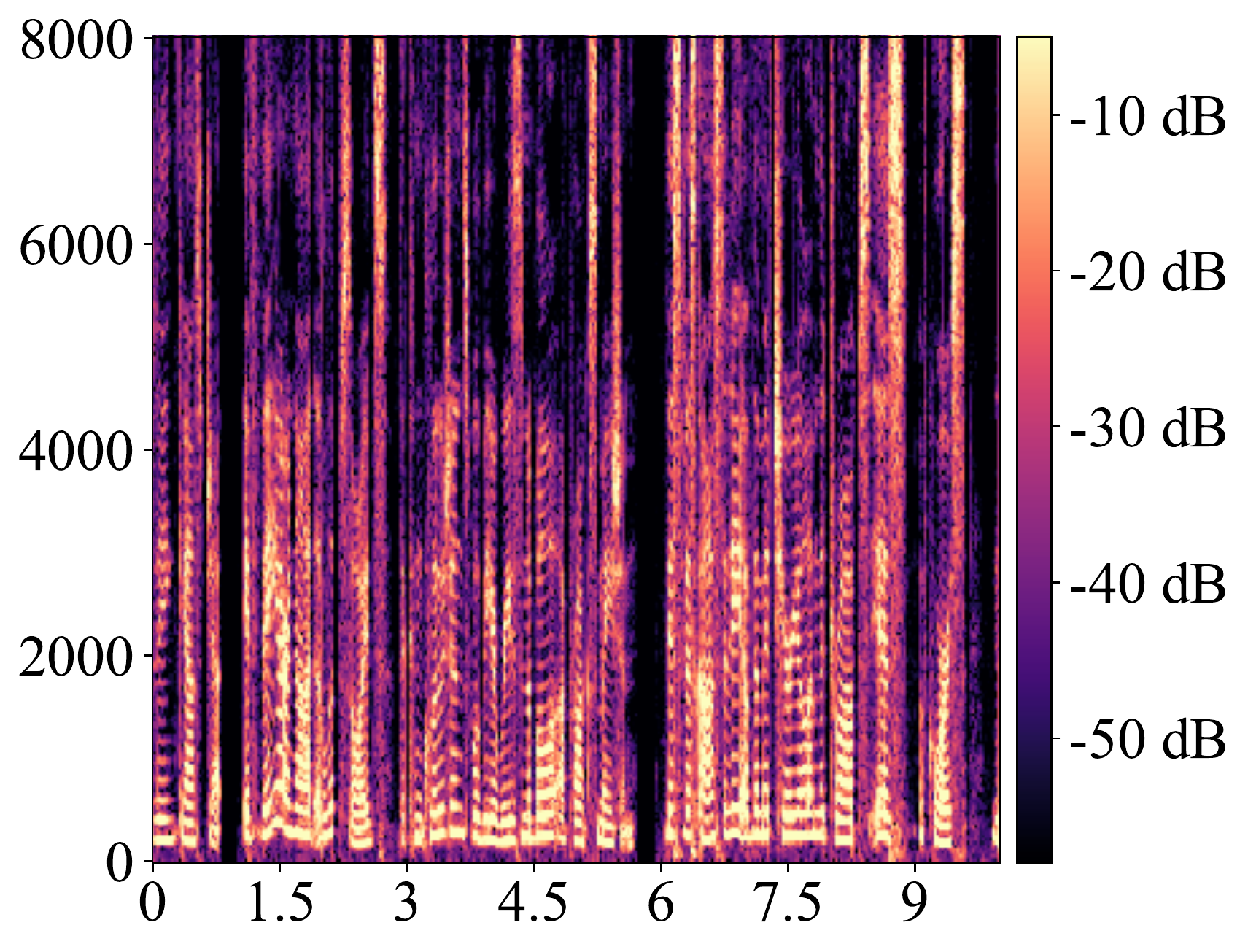}}
\end{minipage}

\begin{minipage}[b]{0.25\textwidth}
  \centering
    \centerline{\footnotesize {\hspace{0.7cm}} (c) Aleatoric-WF}
  \centerline{\includegraphics[width=0.95\linewidth, height=3.7cm]{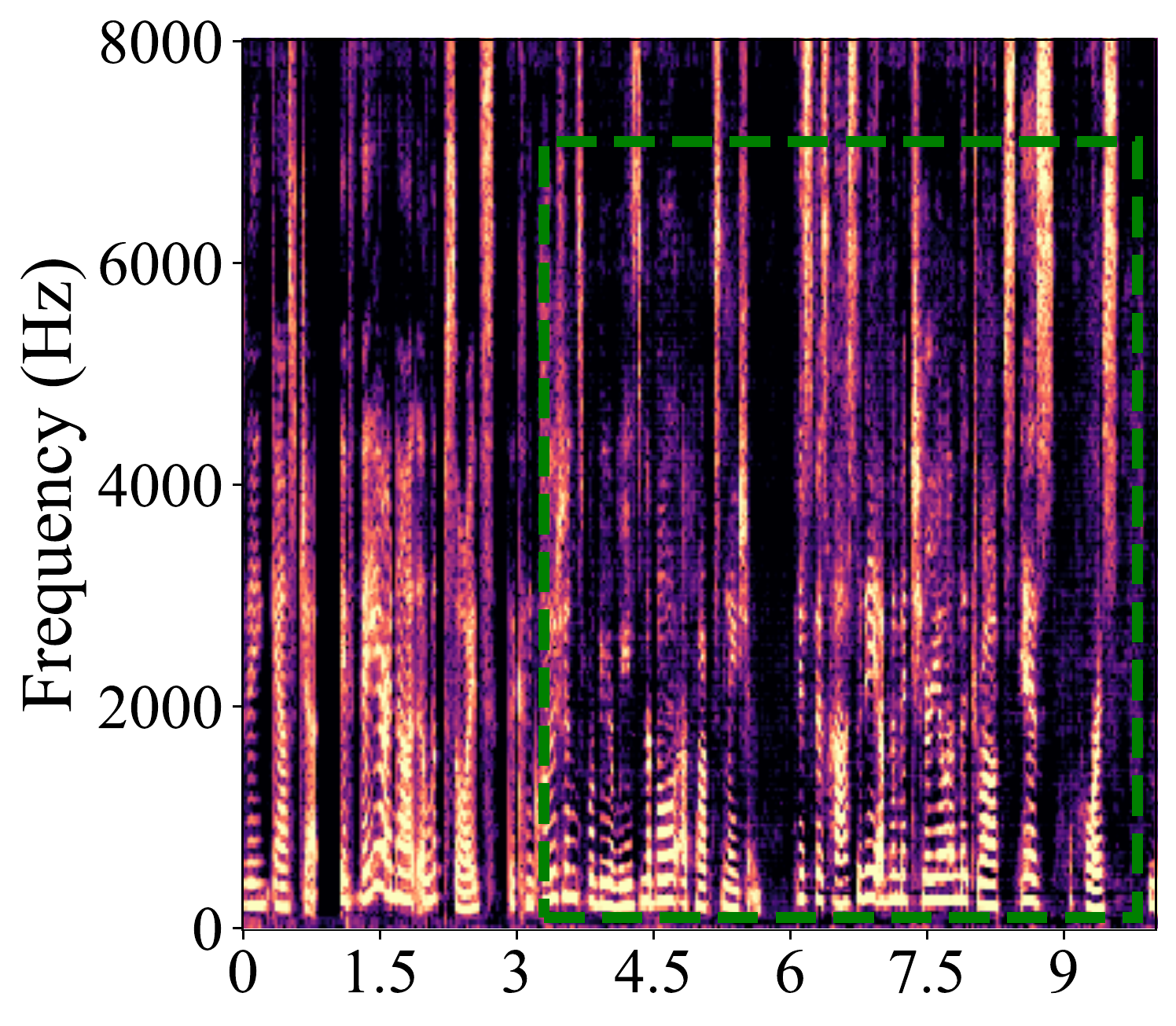}}
\end{minipage}%
\begin{minipage}[b]{0.25\textwidth}
\hspace{-0.5cm}
  \centering
    \centerline{\footnotesize {\hspace{0.8cm}} (d) Error}
  \centerline{\includegraphics[width=0.87\linewidth, height=3.7cm]{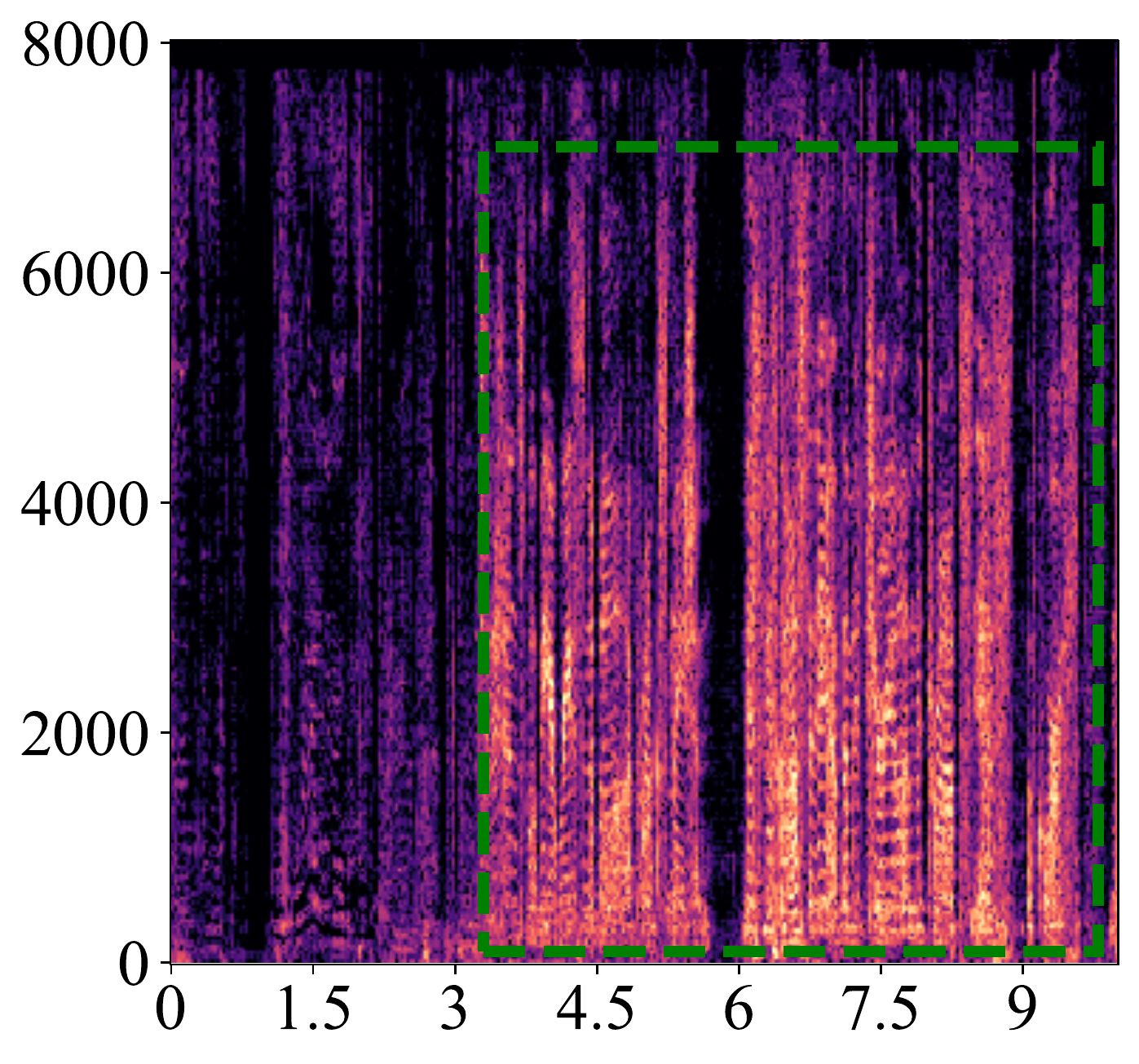}}
\end{minipage}

\begin{minipage}[b]{0.25\textwidth}
  \centering
    \centerline{\footnotesize {\hspace{0.6cm}} (e) Aleatoric uncertainty}
  \centerline{\includegraphics[width=0.95\linewidth, height=3.7cm]{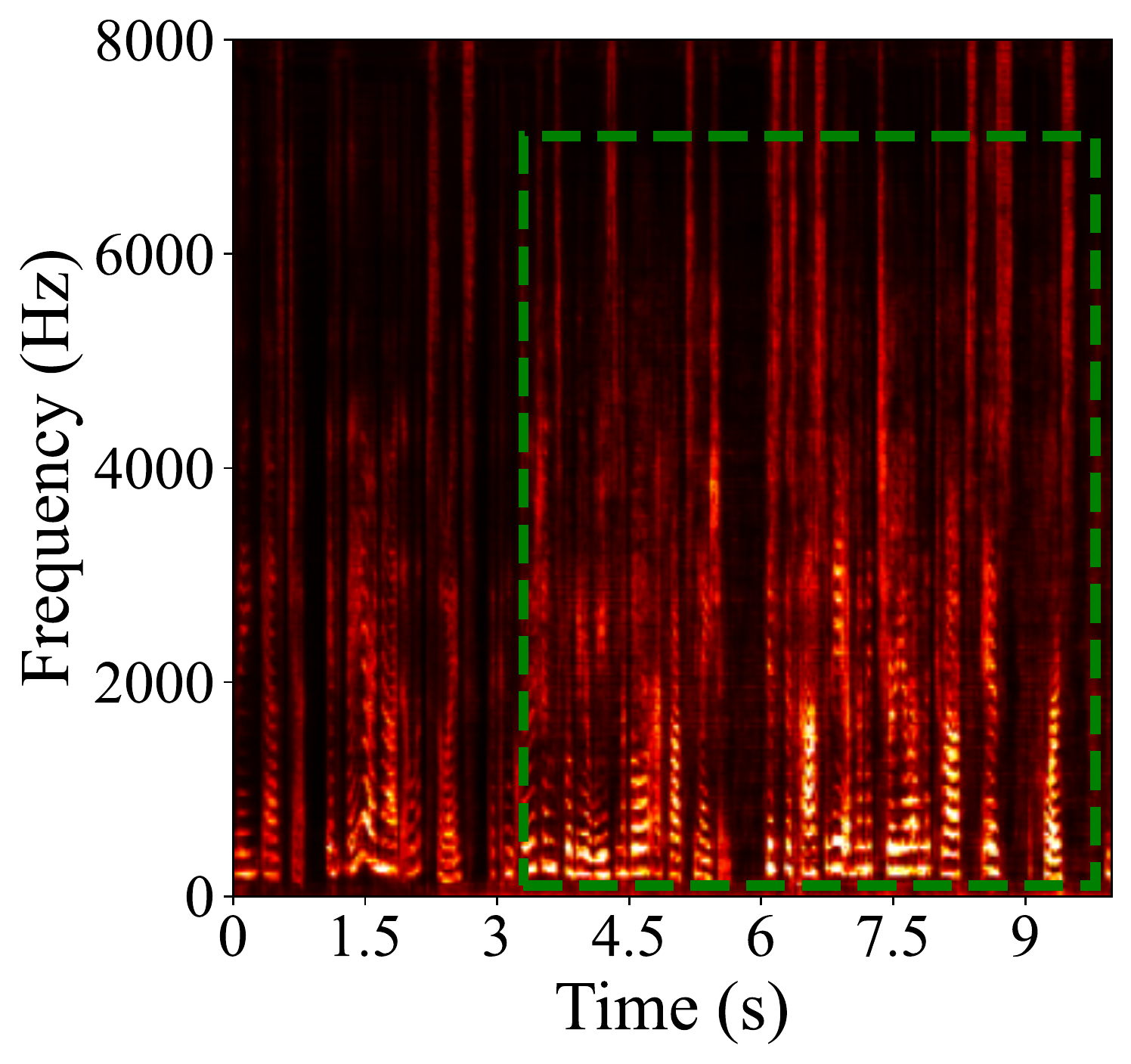}}
\end{minipage}%
\begin{minipage}[b]{0.25\textwidth}
\hspace{-0.5cm}
  \centering
    \centerline{\footnotesize {\hspace{0.8cm}} (f) Aleatoric-AMAP}
  \centerline{\includegraphics[width=0.87\linewidth, height=3.7cm]{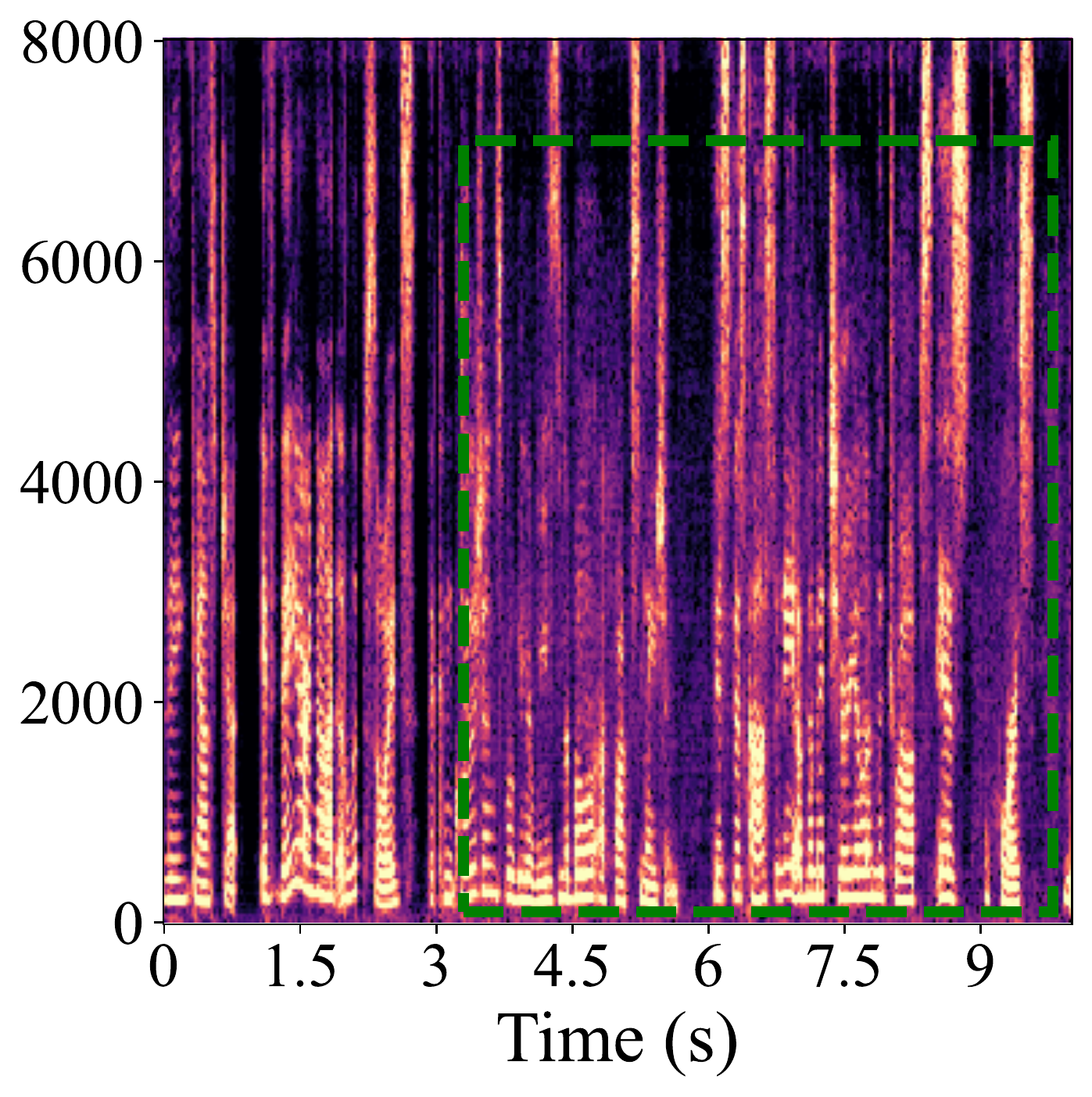}}
\end{minipage}
\caption{Aleatoric uncertainty (shown in (e)) captured by the proposed loss function~(\ref{eq:proposedloss}) for an excerpt from the DNS test dataset. The uncertainty is visualized as a heatmap. The black color indicates low uncertainty, whereas the brighter color indicates higher uncertainty.}
\label{fig:illustrationOfaAleatoricUncertainty}
\vspace{-0.3cm}
\end{figure}

\section{Analysis of uncertainty estimation}
\label{sec:uncertaintyanalysis}
\noindent In this section, we introduce the evaluation metrics for uncertainty and then analyze the captured aleatoric and epistemic uncertainties. Finally, we show that combining two types of uncertainty yields more reliable predictive uncertainty.

\subsection{Uncertainty Evaluation Metrics}
\label{sec:evaluationmetric}
\noindent To evaluate the captured uncertainty, the sparsification plot and the sparsification error are used as evaluation metrics~\cite{ilg2018uncertainty, gustafsson2020evaluating,wannenwetsch2017probflow}. The sparsification plot illustrates the correlation between the uncertainty measure and the true error. The error of a time-frequency bin is defined as the absolute square between the estimated spectral coefficient and the ground-truth. For this plot, the errors in the time-frequency domain are first sorted according to their corresponding uncertainty measures. The residual error should gradually decrease when the time-frequency bins with large uncertainties are removed. This leads to a plot of the \ac{RMSE} versus the fraction of removed time-frequency bins. Normalization is applied to ensure that the plot is initialized at 1. The best ordering of uncertainty measures is determined by ranking the true errors~\cite{ilg2018uncertainty,wannenwetsch2017probflow}. This provides a lower bound of each sparsification plot, denoted as the \emph{oracle} curve, i.e., when the uncertainty estimates and errors are perfectly correlated, the sparsification plot and the oracle curve coincide. The sparsficiation error is computed as the difference between the sparsification plot and the corresponding oracle curve, and the \ac{AUSE} curve provides a single value that enables comparison of different uncertainty modeling techniques. A lower \ac{AUSE} value (i.e., the closer the sparsification plot is to its oracle curve) indicates a more accurate estimate of uncertainty.

\begin{figure}[t]
  \centering
  \centerline{\includegraphics[width=0.75\linewidth]{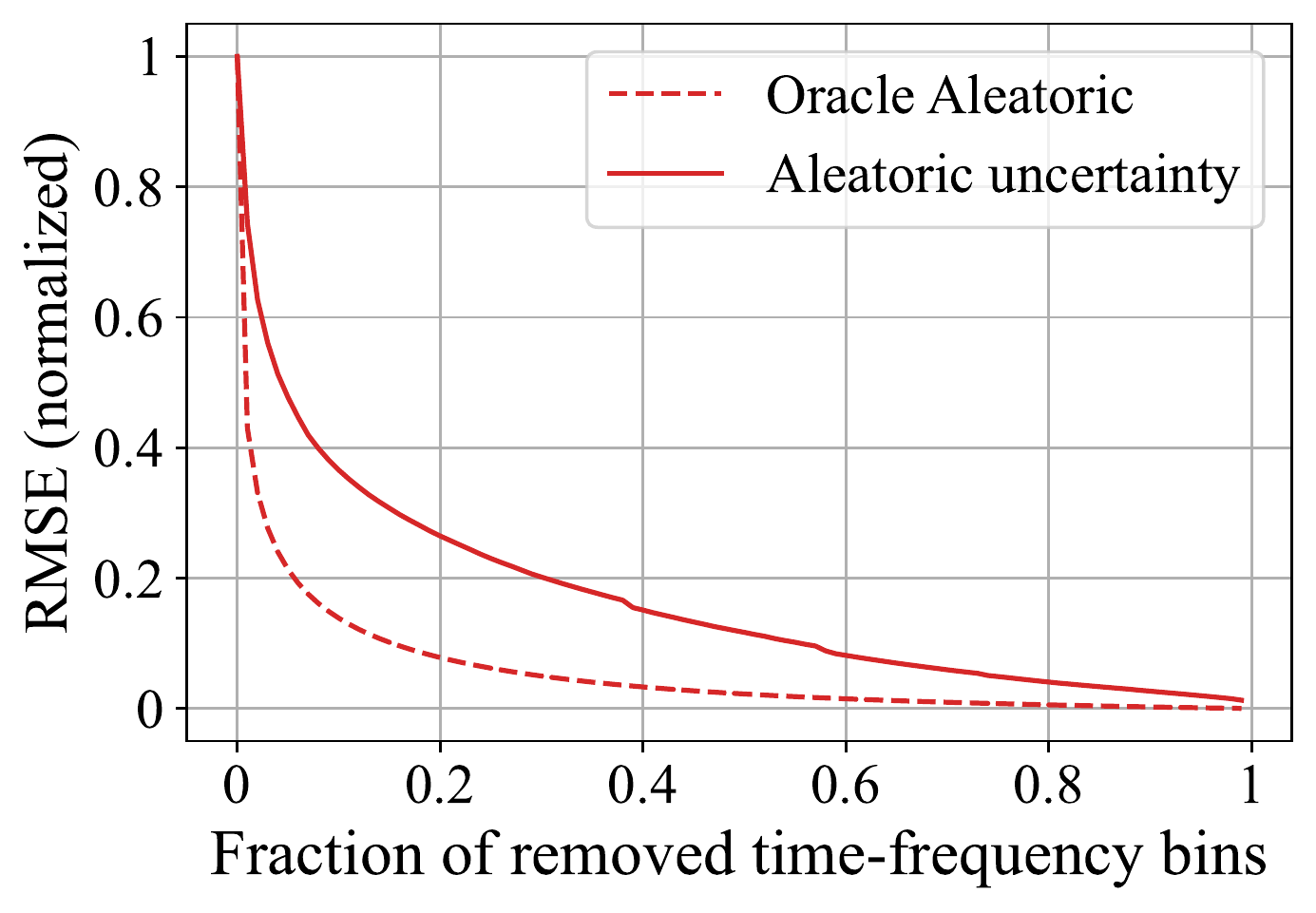}}
  \caption{Sparsification plot of aleatoric uncertainty $\tilde{\lambda}$ evaluated on the DNS test dataset. The dashed line denotes the lower bound of the sparsification plot of aleatoric uncertainty. A smaller distance of the sparsification plot to the oracle curve indicates a more accurate uncertainty estimation.}
\label{fig:sparsification_aleatoric}
\vspace{-0.3cm}
\end{figure}

\begin{figure}[t]
\begin{minipage}[b]{0.25\textwidth}
  \centering
  \centerline{\footnotesize {\hspace{0.8cm}} (a) Estimate (MC dropout)}
  \centerline{\includegraphics[width=0.95\linewidth, height=3.6cm]{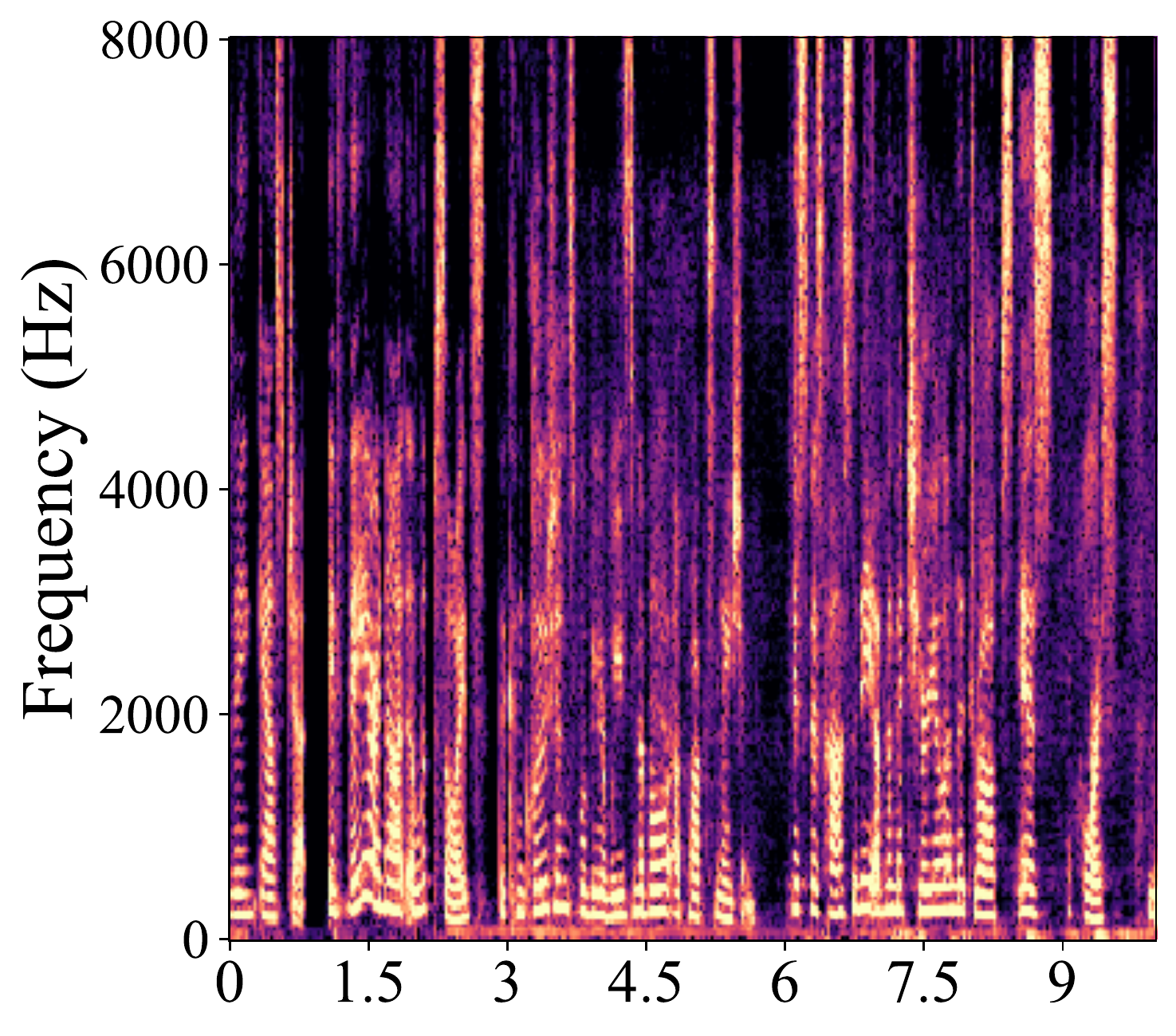}}
\end{minipage}%
\begin{minipage}[b]{.25\textwidth}
  \centering
   \centerline{\footnotesize {\hspace{0.5cm}} (b) Estimate (DE)}
  \centerline{\includegraphics[width=0.9\linewidth, height=3.6cm]{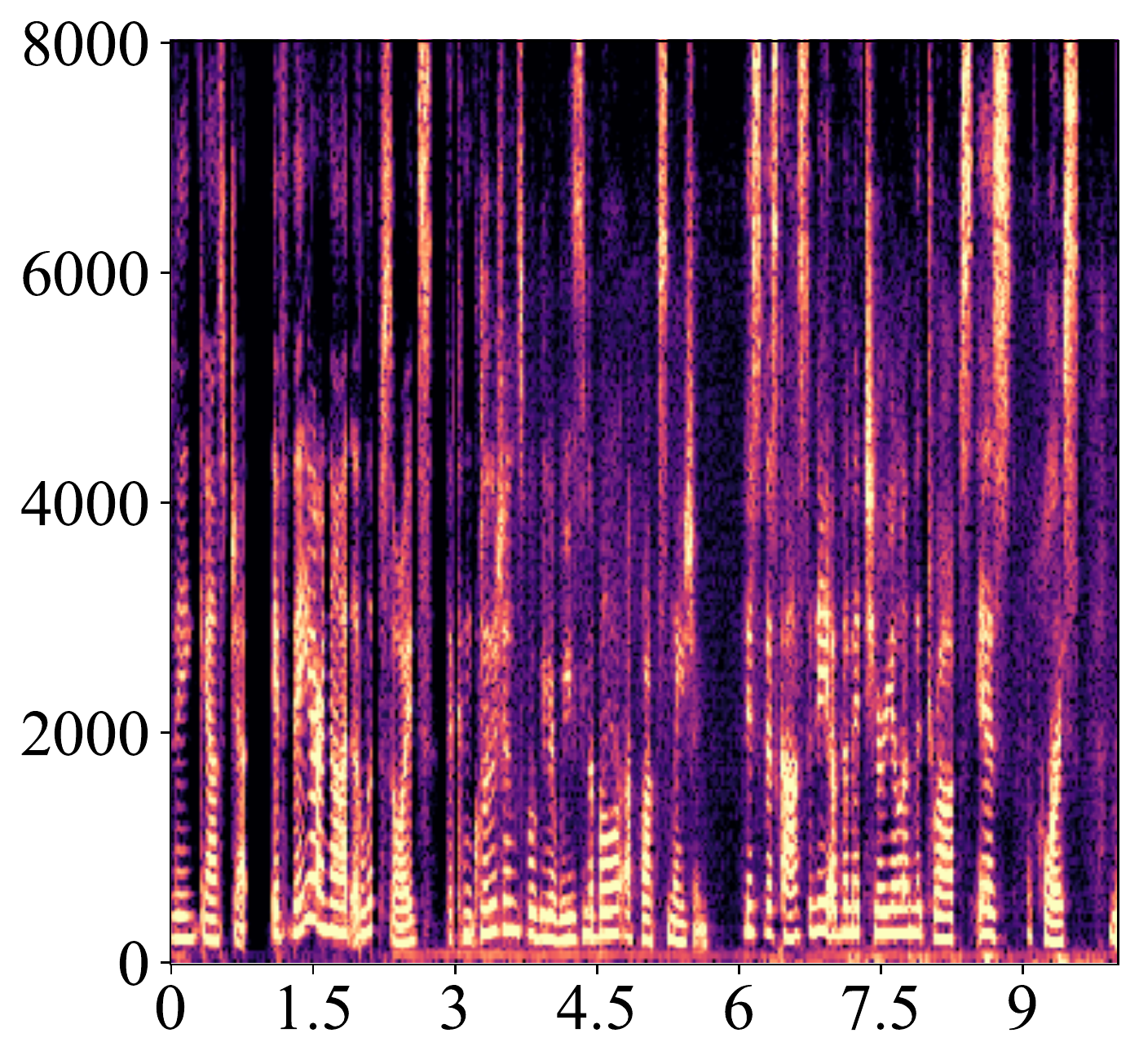}}
\end{minipage}

\begin{minipage}[b]{0.25\textwidth}
  \centering
    \centerline{\footnotesize {\hspace{1cm}} (c) Error (MC dropout)}
  \centerline{\includegraphics[width=0.95\linewidth, height=3.7cm]{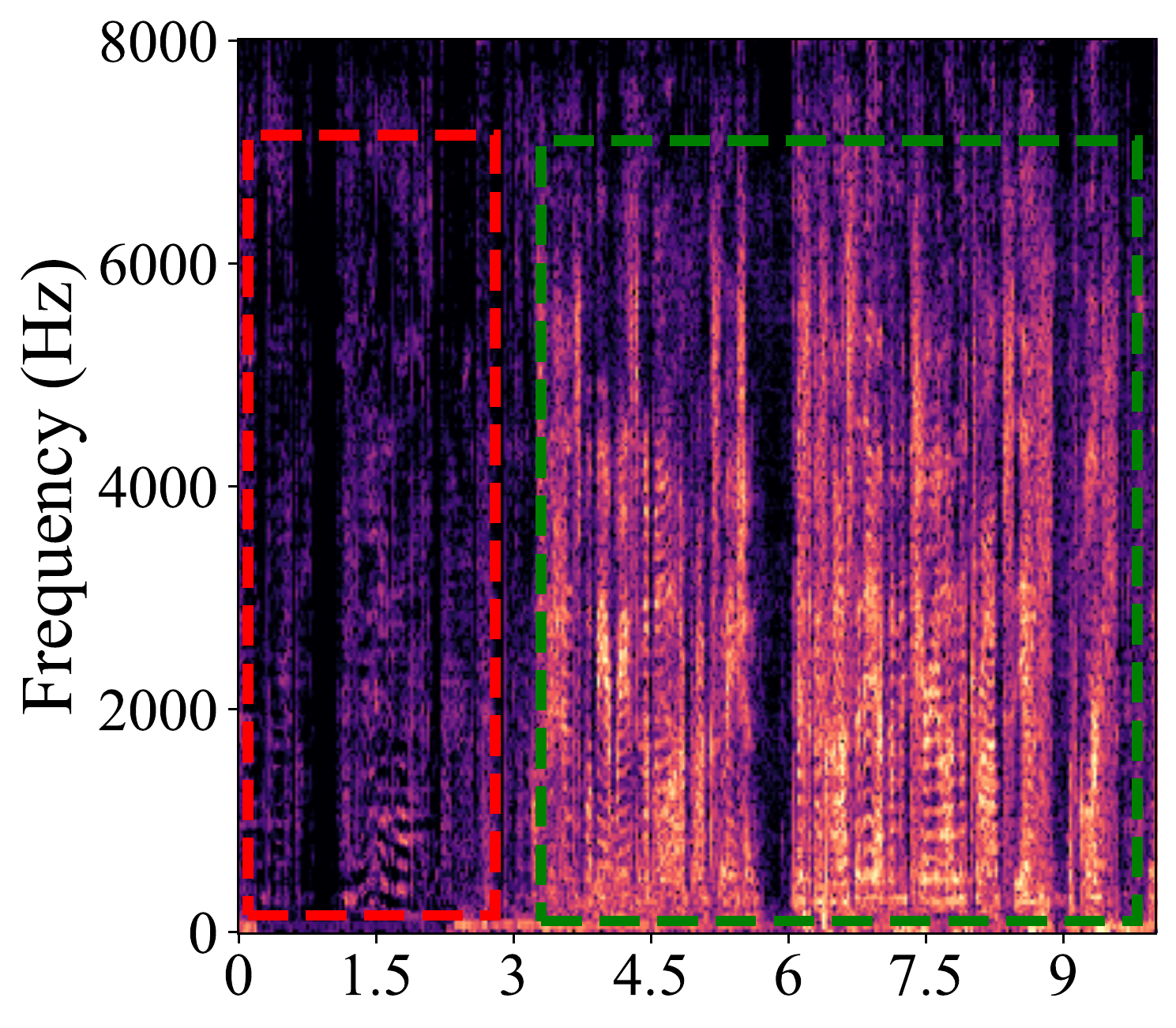}}
\end{minipage}%
\begin{minipage}[b]{0.25\textwidth}
  \centering
    \centerline{\footnotesize {\hspace{0.4cm}} (d) Error (DE)}
  \centerline{\includegraphics[width=0.9\linewidth, height=3.7cm]{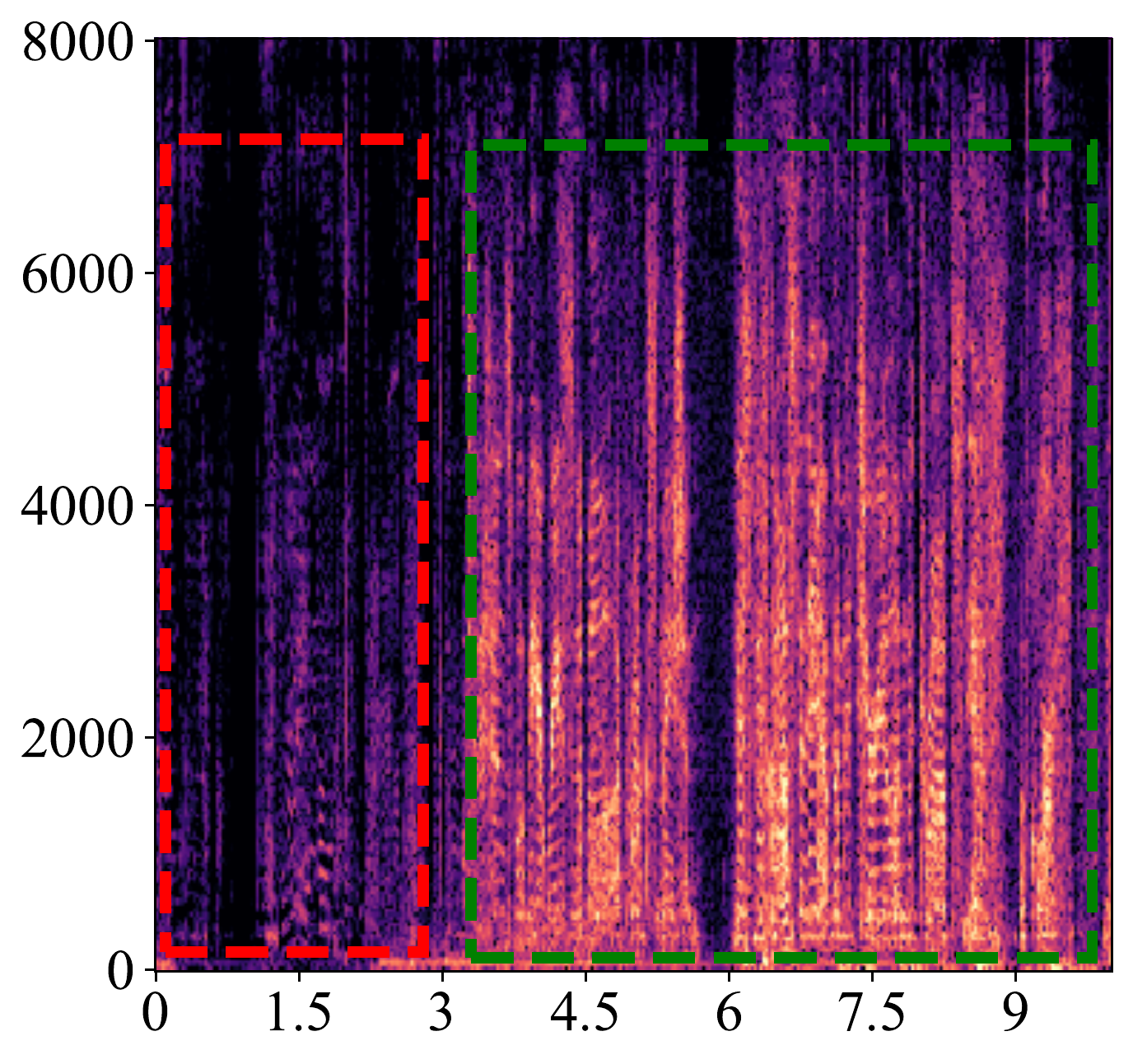}}
\end{minipage}

\begin{minipage}[b]{0.25\textwidth}
  \centering
  \centerline{\footnotesize {\hspace{0.3cm}} (e) Epistemic uncertainty (MC dropout)}
  \centerline{\includegraphics[width=0.95\linewidth, height=3.7cm]{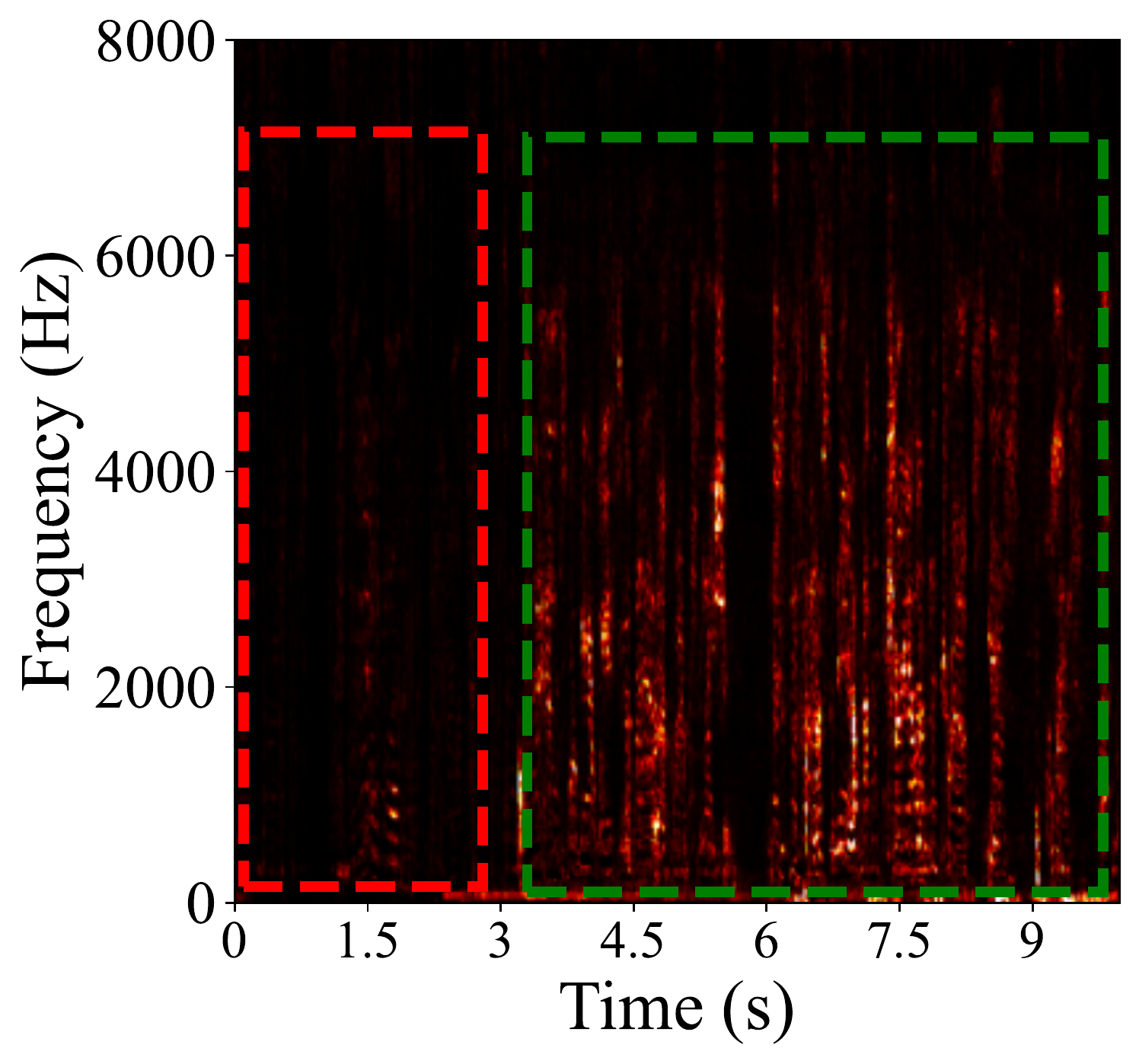}}
\end{minipage}%
\begin{minipage}[b]{0.25\textwidth}
  \centering
  \centerline{\footnotesize {\hspace{0.3cm}} (f) Epistemic uncertainty (DE)}
  \centerline{\includegraphics[width=0.9\linewidth, height=3.7cm]{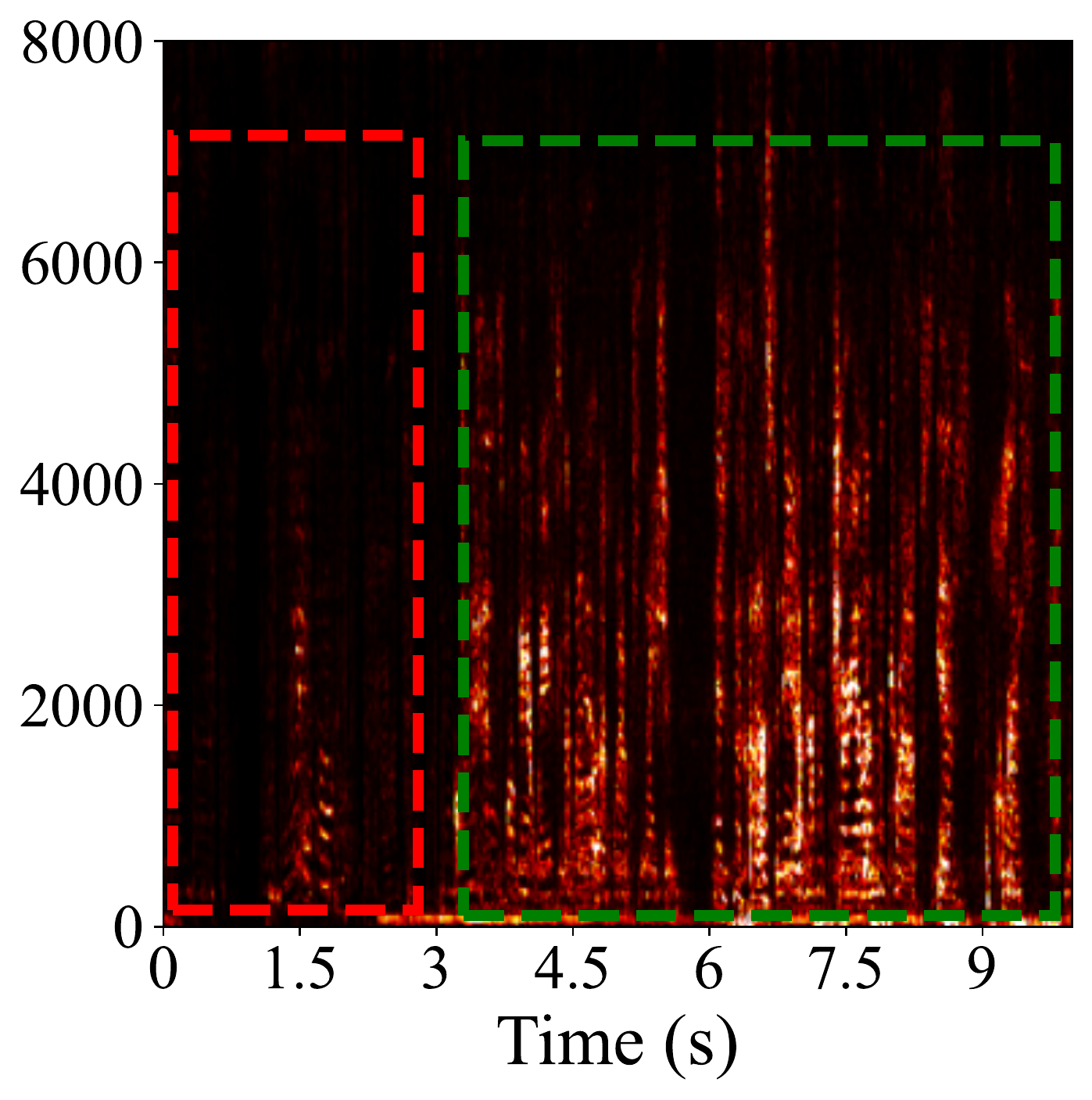}}
\end{minipage}
\caption{The same excerpt as in Fig.~\ref{fig:illustrationOfaAleatoricUncertainty} illustrates the captured epistemic uncertainty obtained by applying Bayesian deep learning methods ($M=16$). \emph{Estimate (MC dropout)} and \emph{Estimate (DE)} represent clean speech estimated using MC dropout and Deep ensembles.}
\label{fig:illustrationEpistemicUncertainty}
\vspace{-0.4cm}
\end{figure}

\begin{figure}[ht!]
  \centering
\begin{minipage}[b]{0.39\textwidth}
  \centering
  \centerline{\includegraphics[width=0.95\linewidth]{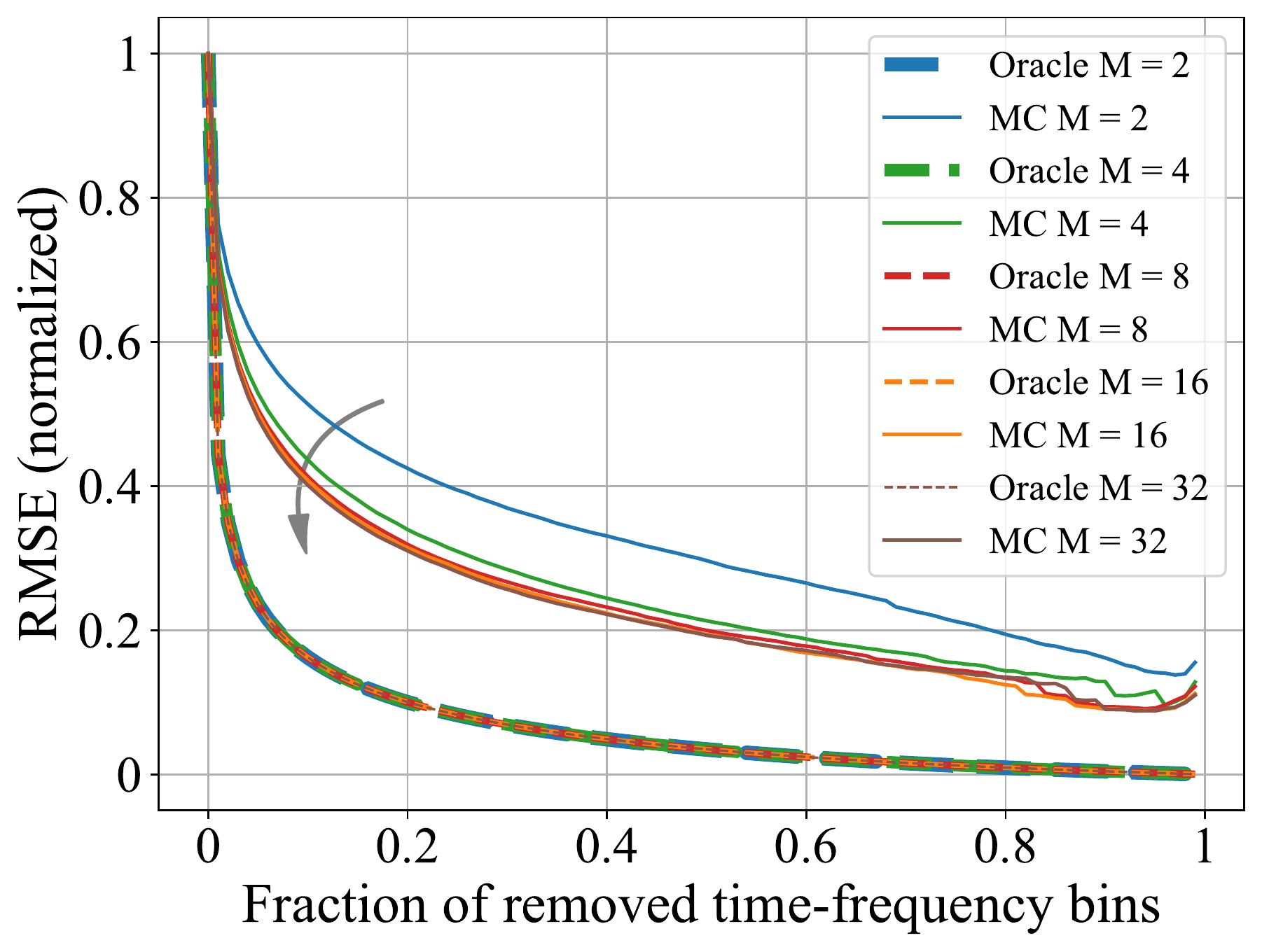}}
\vspace{-0.1cm}
  \centerline{\small (a) Sparsification plot of MC dropout~(MC)}
\end{minipage}

\vspace{0.3cm}

\begin{minipage}[b]{0.39\textwidth}
  \centering
  \centerline{\includegraphics[width=0.95\linewidth]{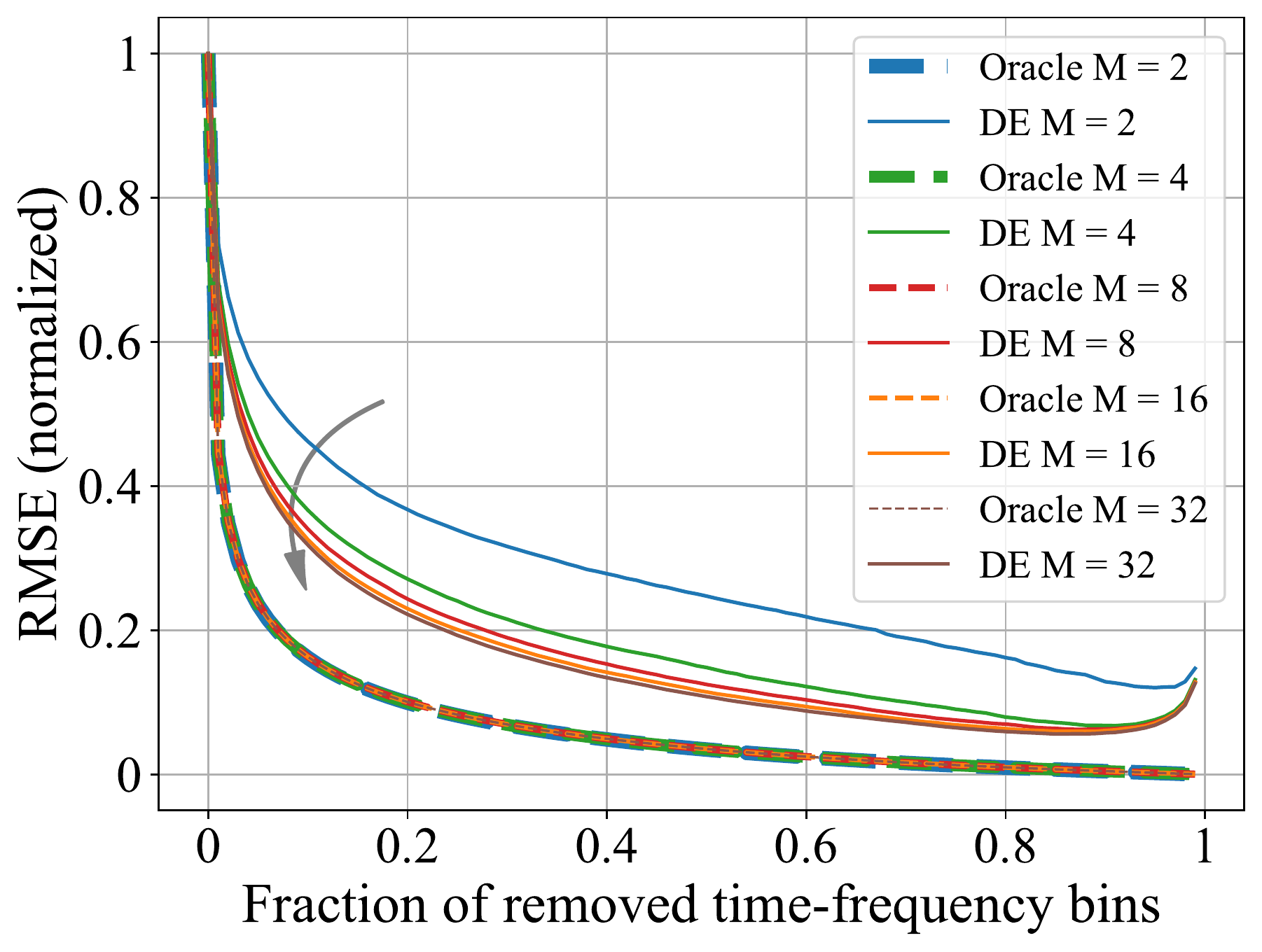}}
\vspace{-0.1cm}
  \centerline{\small (b) Sparsification plot of Deep ensembles~(DE)}
\end{minipage}
\caption{Sparsification plots of epistemic uncertainty $\widetilde{\Sigma}$ for the DNS test dataset. The dashed line denotes the lower bound of the corresponding sparsification plot, denoted as Oracle $M$. A smaller distance of the sparsification plot to the oracle curve indicates a more accurate uncertainty estimation. Note that all oracle curves are visually \emph{overlapping}. }
\label{fig:sparsification_plot_epi}
 \vspace{-0.2cm}
\end{figure}

\begin{figure}[ht!]
  \centering
  \centerline{\includegraphics[width=0.65\linewidth]{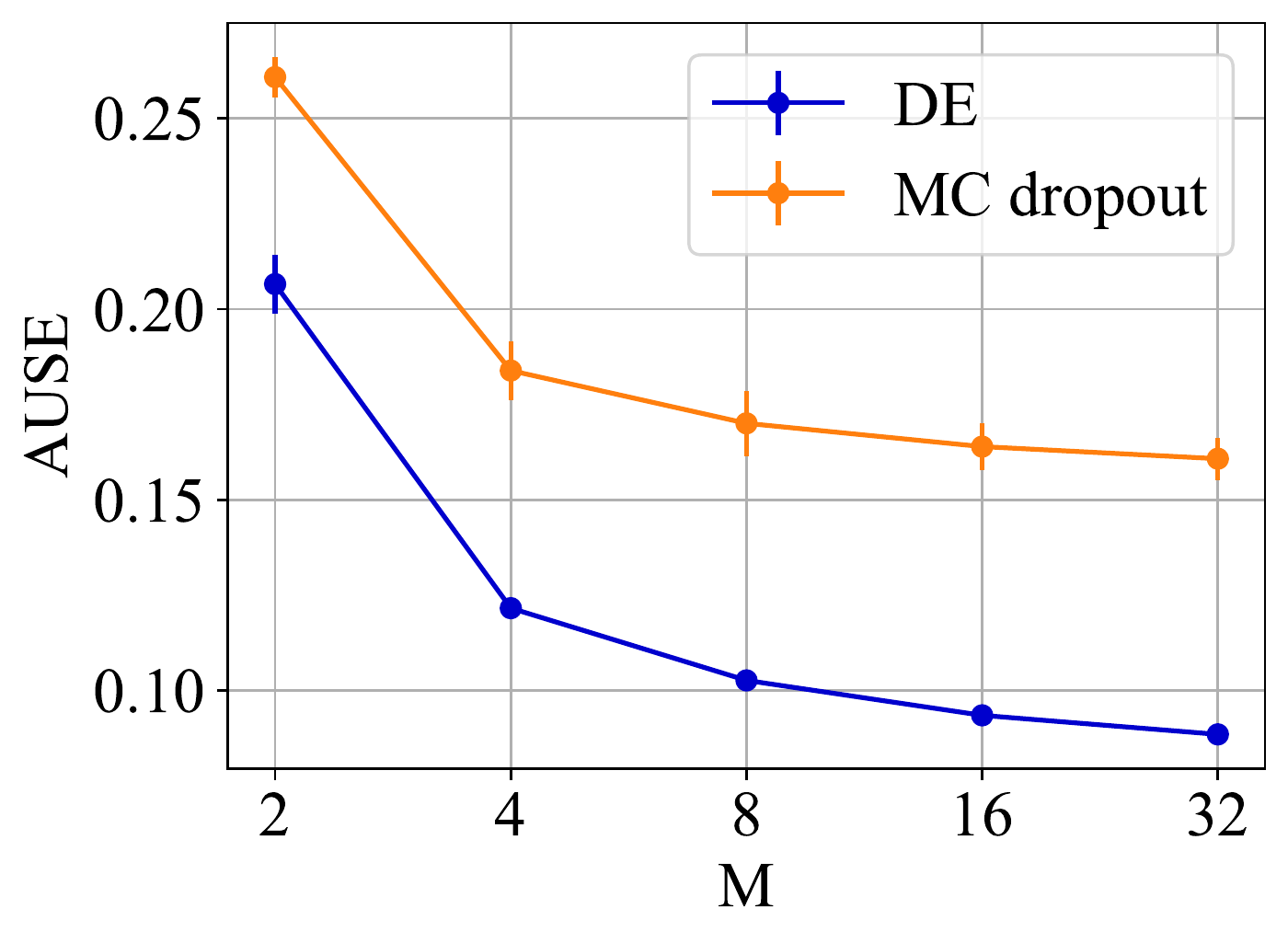}}
    \vspace{-0.2cm}
  \caption{AUSE for the DNS test dataset. AUSE is plotted relative to a different number of forward passes $M$. The markers denote the mean and the vertical bars indicate the standard deviation. Lower values indicate a smaller deviation from the oracle curve, and thus more reliable uncertainty estimation.}
  \vspace{-0.4cm}
\label{fig:sparsification_error_epi}
\end{figure}

\subsection{Analysis of Aleatoric Uncertainty Estimation}
\label{sec:analysisaleatoric}
\noindent In this part, we analyze the captured data-dependent aleatoric uncertainty associated with the Wiener estimate. For this, an audio example from the DNS challenge test set is selected to illustrate the effectiveness of the proposed optimization metric in modeling uncertainty. Aleatoric-WF in Fig.~\ref{fig:illustrationOfaAleatoricUncertainty}~(c) shows the spectrogram of the clean speech obtained by applying the estimated Wiener filter. By computing the absolute square between the clean reference and estimated spectral coefficients, we can obtain the estimation error as depicted in Fig.~\ref{fig:illustrationOfaAleatoricUncertainty}~(d). It can be observed that large errors occur when the speech is heavily disturbed by noise, as in the region marked by the green box, while for inputs with less distortion, such as the first three seconds, the model produces smaller errors. Meanwhile, the proposed loss function enables the estimation of uncertainty associated with the Wiener filter, as shown in Fig.~\ref{fig:illustrationOfaAleatoricUncertainty}~(e), denoted as aleatoric uncertainty. It shows that aleatoric uncertainty prevails in speech presence regions. By relating Fig.~\ref{fig:illustrationOfaAleatoricUncertainty}~(d) to Fig.~\ref{fig:illustrationOfaAleatoricUncertainty}~(e), the model outputs relatively large uncertainty (e.g., the green box-marked part) when large errors are produced. This suggests that the neural network is able to produce reasonable uncertainty estimates when dealing with complex unseen inputs. Furthermore, we can incorporate the estimated uncertainty into clean speech inference, as in~(\ref{eq:approximated_map}), which leads to a clean speech estimate shown in Fig.~\ref{fig:illustrationOfaAleatoricUncertainty}~(f), denoted as Aleatoric-AMAP. It is observed that more speech is preserved than Aleatoric-WF in the highly-uncertain green box-marked region at some cost of noise reduction, i.e., Aleatoric-AMAP leads to less speech distortion with a slight tendency of retaining more noise. The reason for this is that with reliable uncertainty estimates, Aleatoric-AMAP can increase the estimator's value in~\eqref{eq:approximated_map} under high uncertainty~(as the AMAP estimator’s value is positively correlated with the uncertainty estimate when other terms are fixed), thus causing less target attenuation.

Besides the qualitative analysis, we can associate the captured uncertainty with the corresponding prediction errors on the time-frequency bin scale and use sparsification plots to analyze the reliability of the uncertainty estimates. The sparsification plot shown in Fig.~\ref{fig:sparsification_aleatoric} is computed based on all audio samples in the DNS reverb-free test dataset. We observe a rapid decrease at the beginning in Fig.~\ref{fig:sparsification_aleatoric}, implying that large errors come with large uncertainty estimates. By removing 20 percent of time-frequency bins with high uncertainty (i.e., 0.2 in the horizontal axis), the~\ac{RMSE} value drops by around two-thirds. Thus, the monotonically decreasing sparsification plot in Fig.~\ref{fig:sparsification_aleatoric} again suggests that the predicted aleatoric uncertainty measurement is closely related to the estimation error. 

\subsection{Analysis of Epistemic Uncertainty Estimation}
\label{sec:epistemicAnalysis}

\noindent Next, we ignore aleatoric uncertainty and analyze separately epistemic uncertainty in the model parameters. For this, the neural networks are trained to perform only point estimation, i.e., trained with the loss function~(\ref{eq:mse}). An ensemble of models is collected by applying Deep ensembles or MC dropout to approximate the predictive mean and variance.

In Fig.~\ref{fig:illustrationEpistemicUncertainty}, we present the same audio example as in Fig.~\ref{fig:illustrationOfaAleatoricUncertainty} to illustrate the uncertainty measures based on MC dropout and Deep ensembles. MC dropout and Deep ensembles provide the clean speech estimates as shown in the first row of Fig.~\ref{fig:illustrationEpistemicUncertainty}. The estimation error for each method is obtained similarly by calculating the absolute square between the estimated and clean spectral coefficients, shown in the second row. As can be observed, both methods produce large errors as well as associated large uncertainties when the signal is heavily corrupted by noise, i.e., the green box-marked region. While the noise corruption is less severe, i.e., the region marked with a red box, the model generates low prediction errors and also a relatively low level of uncertainty. From the visual analysis, the uncertainty generated by Deep ensembles is more correlated with the error, while MC dropout appears to underestimate the uncertainty of incorrect predictions. To objectively assess the reliability of uncertainty measures, we also utilize the sparsification plots and the sparsification errors, as illustrated in Fig.~\ref{fig:sparsification_plot_epi} and Fig.~\ref{fig:sparsification_error_epi} respectively.

In Fig.~\ref{fig:sparsification_plot_epi}, we show the sparsification plots of Deep ensembles and MC dropout for a different number of forward passes $M \in \{2,4,8,16,32\}$. It can be observed that both MC dropout and Deep ensembles yield decreasing sparsification plots, suggesting that they produce accurate uncertainties that correlate well with the estimation errors. It also shows that a large $M$ leads to a sparsification plot closer to its corresponding oracle curve, i.e., improves the performance of the uncertainty estimation, and this improvement becomes saturated when $M$ is sufficiently large, e.g., from $M=16$ to $M=32$.

To comprehensively compare MC dropout and Deep ensembles in terms of uncertainty modeling, \ac{AUSE} is plotted as a function of different numbers of forward passes $M$. Multiple models for each $M$ are used to provide mean and standard deviation to account for variations resulting from random factors in training. 16 MC dropout models are trained and used to compute the mean of \ac{AUSE} and its standard deviation for each possible $M$. For Deep ensembles, 16 disjoint sets of $M$ models are randomly selected from the $33$ trained models to compute the mean and standard deviation of \ac{AUSE}. The AUSE plot in~Fig.~\ref{fig:sparsification_error_epi} provides an alternative and more informative evaluation than a single sparsification plot. It indicates that Deep ensembles generally produce more accurate uncertainty than MC dropout, which may fail to produce reliable uncertainties for some erroneous predictions. This coincides with our visual observation in the green box-marked region in Fig.~\ref{fig:illustrationEpistemicUncertainty}. 

\begin{figure}[tbp]
  \centering
  \centerline{\includegraphics[width=0.75\linewidth]{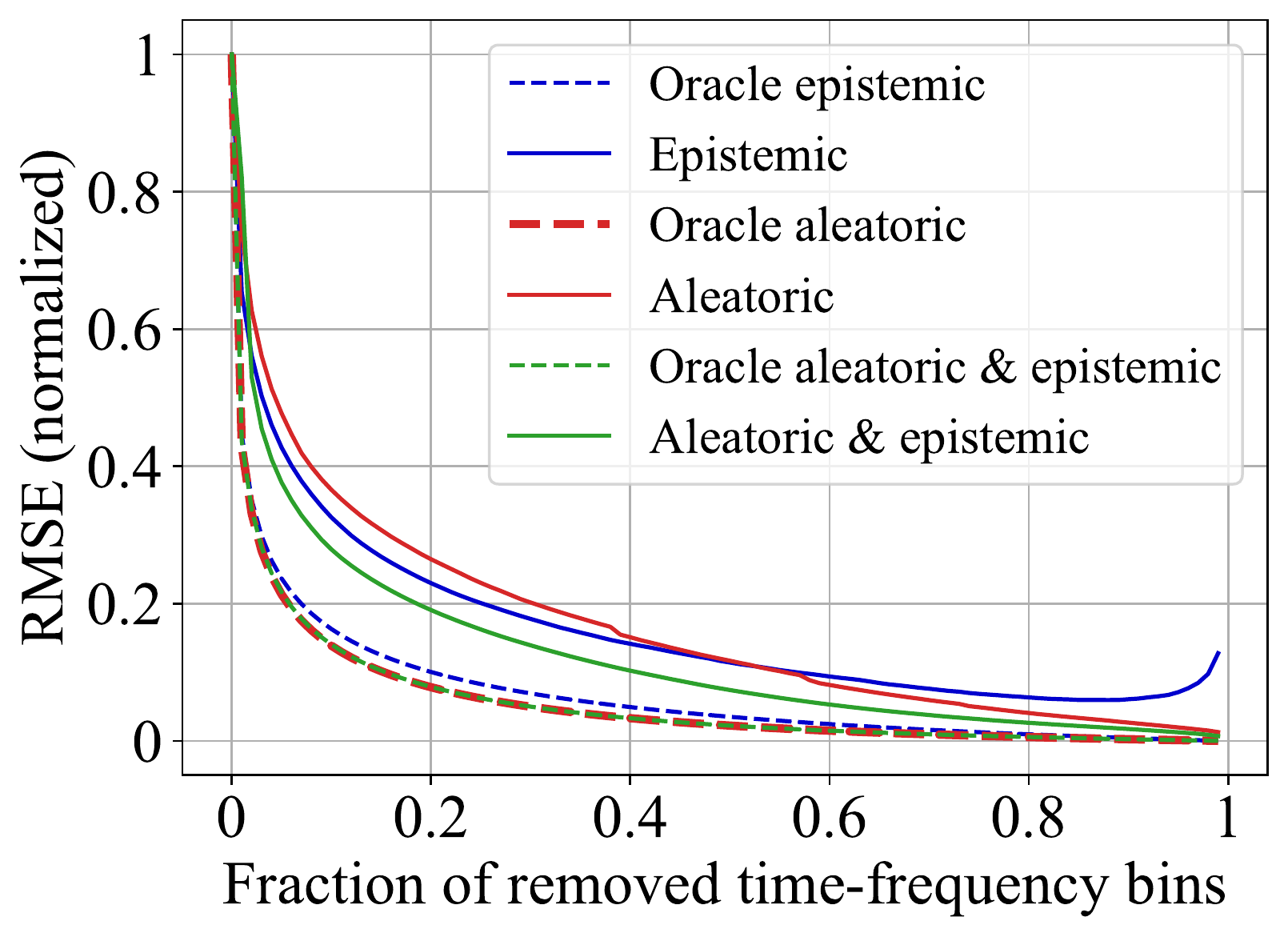}}
  \caption{Sparsification plots of aleatoric~$\tilde{\lambda}$, epistemic~$\widetilde{\Sigma}$ , and overall predictive uncertainty $\widehat{\Sigma}$ (i.e., aleatoric \& epistemic) on the DNS test dataset. Note that \emph{Oracle aleatoric} and \emph{Oracle aleatoric \& epistemic} overlap.}
\label{fig:sparsification_plot_predictive}
\end{figure}

\begin{table}[tbp]
\caption{AUSE values of \emph{Aleatoric}, \emph{Epistemic}, and \emph{Aleatoric \& epistemic} in Fig.~\ref{fig:sparsification_plot_predictive}.}
\vspace{-0.1cm}
\begin{center}
\resizebox{0.9\columnwidth}{!}{\begin{tabular}{|c|c|c|c|}
\hline
 & Aleatoric & Epistemic & Aleatoric \& epistemic\\
\hline
AUSE & 0.110 & 0.094 & 0.067 \\
\hline
\end{tabular}}
\end{center}
\label{AUSE43uncer}
\vspace{-0.4cm}
\end{table}

\subsection{Prediction Uncertainty Combining Aleatoric and Epistemic Uncertainties}
\label{analysispredictiveuncertainty}
\noindent In this part, we investigate the overall prediction uncertainty obtained by combining aleatoric uncertainty and epistemic uncertainty as in~\eqref{mean_wf_var_pre}. To obtain the overall prediction uncertainty, we use an ensemble of models trained with the optimization metric~\eqref{eq:proposedloss} such that both aleatoric and epistemic uncertainty are captured. It has been shown in Section~\ref{sec:epistemicAnalysis} that Deep ensembles yield more accurate epistemic uncertainty than MC dropout and, therefore, are selected for the estimation of the overall predictive uncertainty. Although a larger number of models $M$ could potentially improve the mean and variance estimation, we restrict $M$ to 16 as further improvements become subtle while the computation time increases considerably.

In Fig.~\ref{fig:sparsification_plot_predictive}, we use sparsification plots to analyze the quality of prediction uncertainty estimates combining aleatoric and epistemic uncertainties. The corresponding AUSE values are provided in Table~\ref{AUSE43uncer}. The plot illustrates that the overall predictive uncertainty estimates correlate stronger with the estimation error than either of the two uncertainties alone. This suggests that two sources of uncertainty may complement each other and combining both leads to more reliable uncertainty estimates. For example, Deep ensembles do not seem to capture sufficient uncertainty for less distorted input~(e.g., first three seconds) as shown in Fig.~\ref{fig:illustrationEpistemicUncertainty}, while aleatoric uncertainty shown in Fig.~\ref{fig:illustrationOfaAleatoricUncertainty} could be able to compensate for this shortcoming.

\section{Influence of modeling uncertainty for speech enhancement performance}
\label{sec:results}

\begin{table}[tbp]
\caption{Evaluation results on the DNS test dataset. All results are stated as mean $\pm$ 95\%-confidence interval. \emph{Unc.} stands for \emph{Uncertainty}.}
\begin{center}
\resizebox{1\columnwidth}{!}{
\begin{tabular}{|c||c||c|c|c|c|}

\hline
 & Unc. & PESQ & ESTOI & SI-SDR\\
\hline
Noisy (DNS) & -& 1.58 $\pm$ 0.07 & 0.81 $\pm$ 0.02 & 9.07 $\pm$ 0.89\\
\hline
Baseline WF &  \xmark & 2.48 $\pm$ 0.10 & 0.90 $\pm$ 0.01 & 16.84 $\pm$ 0.74\\
Baseline SI-SDR & \xmark & 2.63 $\pm$ 0.10 & 0.91 $\pm$ 0.01 & 17.49 $\pm$ 0.78\\
\hline
MC dropout & \cmark
& 2.53 $\pm$ 0.10  & 0.90 $\pm$ 0.01  & 16.88 $\pm$ 0.74  \\
Deep ensembles & \cmark
 & 2.66 $\pm$ 0.10  & 0.91 $\pm$ 0.01  & 17.16 $\pm$ 0.73  \\
\hline
Aleatoric-WF & \cmark &   2.62 $\pm$ 0.11 &  0.91 $\pm$ 0.01 & 17.54 $\pm$ 0.78\\
Aleatoric-MAP &\cmark & 2.69 $\pm$ 0.10 & 0.91 $\pm$ 0.01 & 17.54 $\pm$ 0.78\\
\hline
DE-Aleatoric-WF & \cmark & 2.77 $\pm$ 0.11 & 0.92 $\pm$ 0.01 & 17.88 $\pm$ 0.78 \\

DE-Aleatoric-AMAP & \cmark & 2.83 $\pm$ 0.10 & 0.92 $\pm$ 0.01 & 17.90 $\pm$ 0.78\\
\hline
\end{tabular}
}
\end{center}
  \label{dns}
  \vspace{-0.4cm}
\end{table}

\noindent In this section, we show how modeling different sources of uncertainty affects the performance of speech enhancement. To evaluate the speech enhancement performance, we employ \ac{PESQ}~\cite{pesq} to measure speech quality, \ac{ESTOI}~\cite{estoi} to measure speech intelligibility, and \ac{SI-SDR} to account for both noise reduction and speech distortion.

\begin{figure}[tp]
\vspace{-0.1cm}
     \centering
    \begin{minipage}[b]{0.45\textwidth}
      \centering
      \centerline{\includegraphics[width=0.86\linewidth, height=8cm]{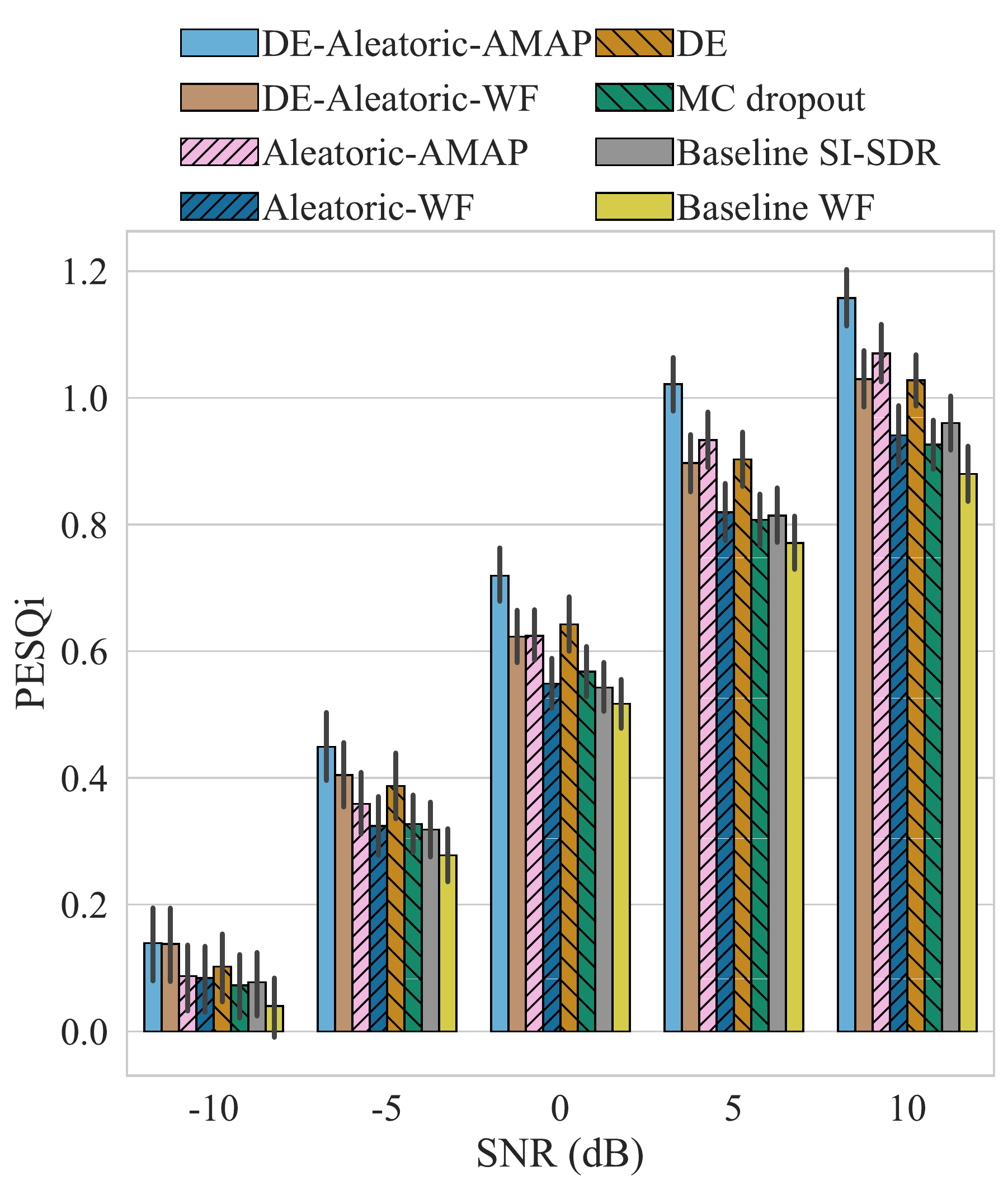}}
    \end{minipage}
    \begin{minipage}[b]{0.45\textwidth}
      \centering
      \centerline{\includegraphics[width=0.87\linewidth, height=5.5cm]{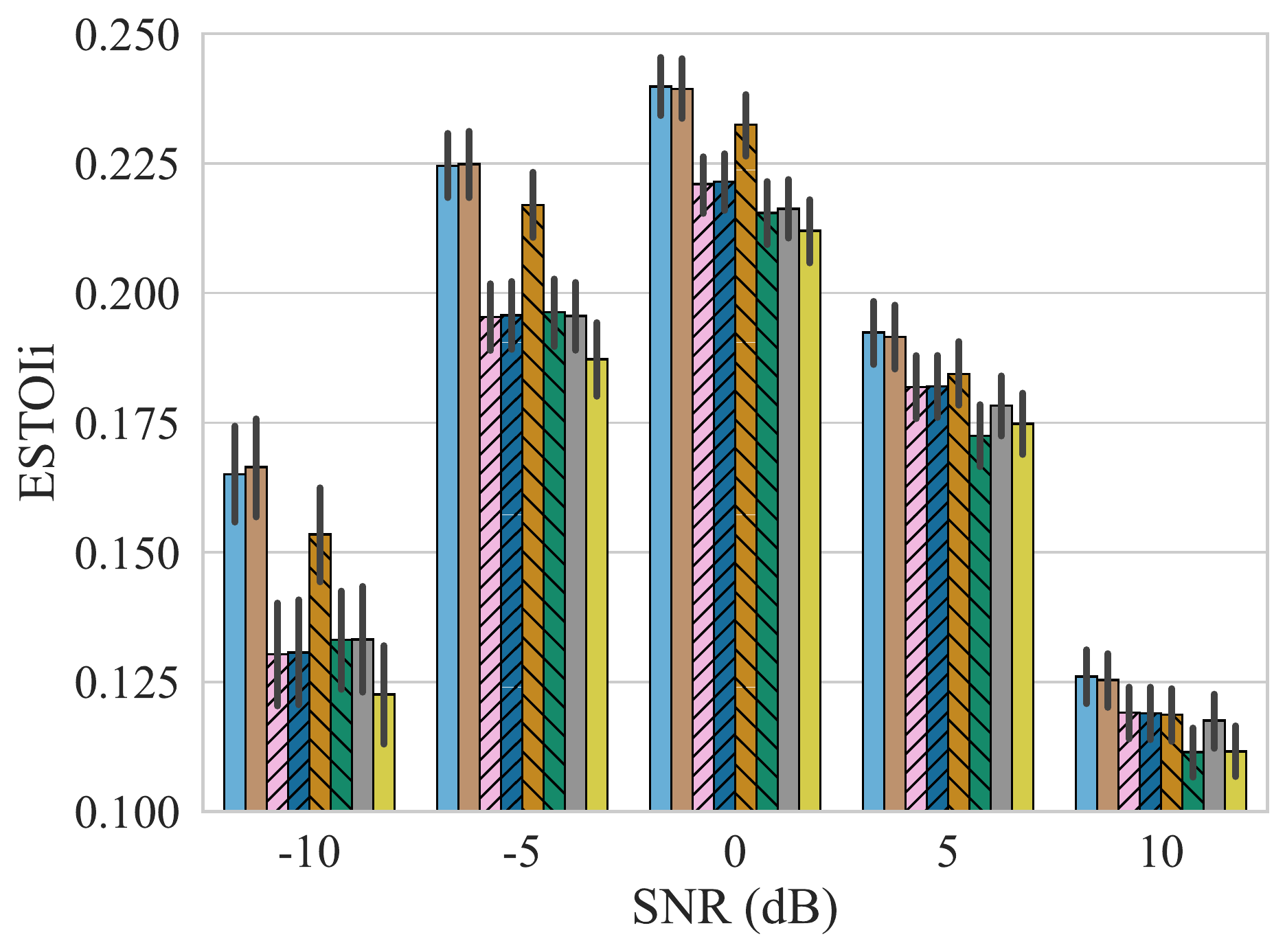}}
    \end{minipage}
    \begin{minipage}[b]{0.45\textwidth}
      \centering
      \centerline{\includegraphics[width=0.85\linewidth, height=5.5cm]{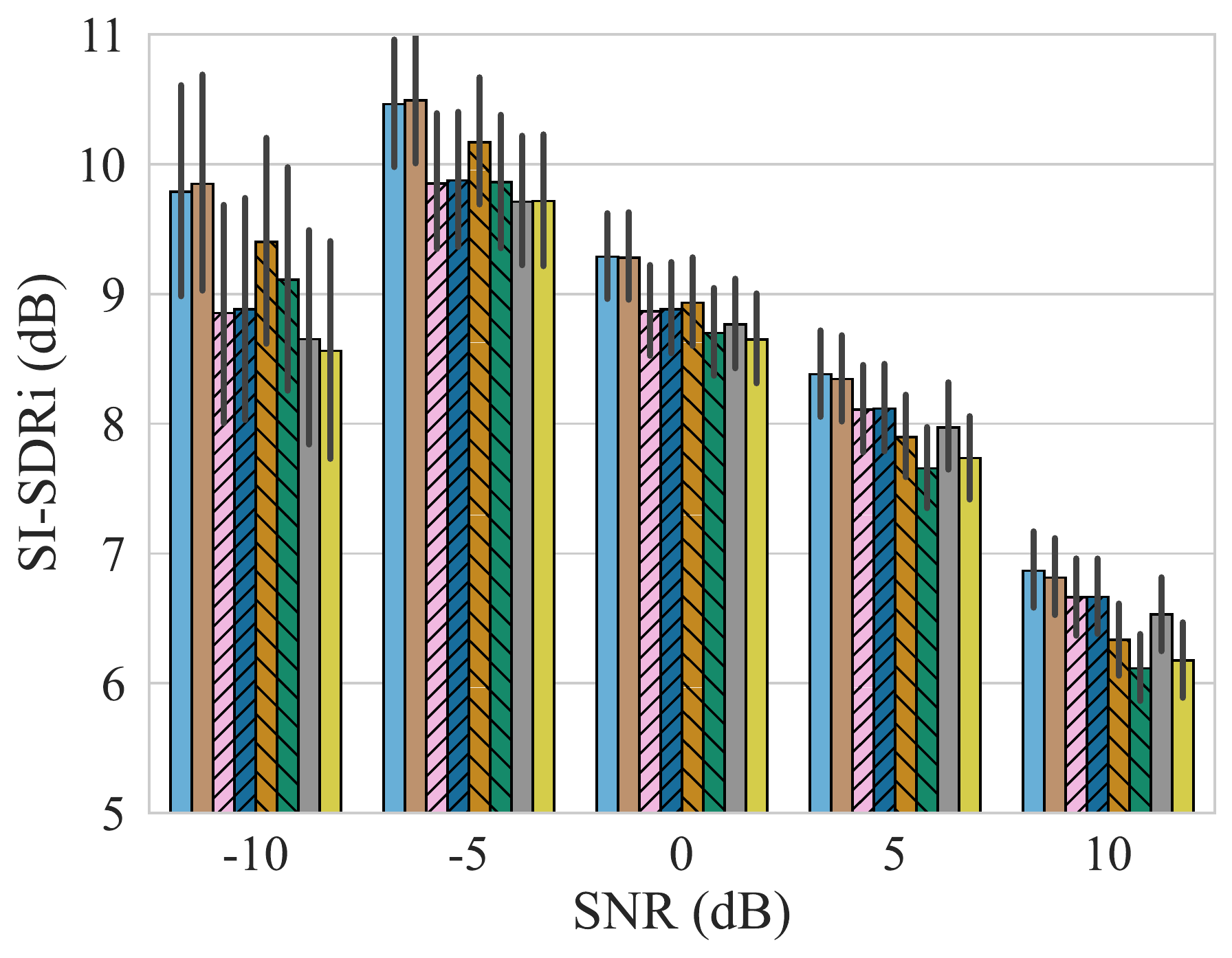}}
    \end{minipage}
    \vspace{-0.1cm}
    \caption{Performance improvement on the dataset with speech from WSJ0 and noise from CHiME3. PESQi denotes PESQ improvement with respect to noisy mixtures. ESTOIi and SI-SDRi are defined similarly. Markers and vertical bars indicate the mean and 95\%~confidence interval.}
    \label{fig:wsjnchime}
\vspace{-0.5cm}
\end{figure}

To show the impact of modeling aleatoric uncertainty on speech enhancement performance, we compare the performance of the model trained with the proposed loss function~\eqref{eq:proposedloss} with that of Baseline WF and Baseline SI-SDR. The proposed method enables speech estimation via either the Wiener filter, which implicitly takes uncertainty into account during the training process, or the approximated \ac{MAP} filter, which explicitly includes uncertainty to estimate speech, denoted as Aleatoric-WF and Aleatoric-AMAP respectively. Table~\ref{dns} shows the average evaluation results on the DNS synthetic non-reverb test set. Aleatoric-WF shows improvements in \ac{PESQ}, \ac{ESTOI}, and \ac{SI-SDR} compared to the Baseline WF, indicating the benefit of weighting Wiener estimates with uncertainty during training. Further \ac{PESQ} improvements over both Baseline WF and Baseline SI-SDR can be observed when explicitly incorporating uncertainty into clean speech estimation, that is, Aleatoric-AMAP. This demonstrates the advantage of modeling uncertainty associated with the Wiener estimate rather than directly estimating optimal points. When evaluated on another dataset with speech from WSJ and noise from CHiME3, the performance gap between Aleatoric-AMAP and the baselines in terms of \ac{PESQ} is further increased, as shown in Fig.~\ref{fig:wsjnchime}, indicating that the model that takes uncertainty into account has improved generalization capacities for speech enhancement. This can be attributed to the nonlinear estimation characteristics of the uncertainty-based AMAP estimator with respect to noisy inputs and the resulting better speech preservation properties. We observe larger improvements over the baselines at high \acp{SNR}, which might be explained by the fact that, at high \acp{SNR}, speech quality (and thus \ac{PESQ}) is mainly affected by speech distortions, while at low \acp{SNR} the main factor is residual noise. Overall, these evaluation results demonstrate the notable benefits of modeling aleatoric uncertainty in the algorithm.

To show the impact of modeling epistemic uncertainty on speech enhancement performance, we compare the performance of Deep ensembles and MC dropout with Baseline WF. We again restrict $M$ to $16$ as in Section~\ref{analysispredictiveuncertainty}. MC dropout performs comparably to Baseline WF on the DNS test set, while a larger improvement can be observed when using Deep ensembles. This improvement is even more pronounced in \ac{PESQ}. Similarly, the results on the second test set are shown in Fig.~\ref{fig:wsjnchime}, where Deep ensembles and MC dropout improve over Baseline WF in terms of \ac{PESQ} for all considered \acp{SNR} and provide higher ESTOI scores, especially at low \acp{SNR}. We observe that Deep ensembles not only provide more accurate uncertainty estimates than MC dropout but also lead to a better speech enhancement performance. A possible explanation is that while MC dropout only captures local uncertainty around a single mode, Deep ensembles trained with different initialization points are capable of exploring multiple modes in the function space to account for training data, see, e.g.,~\cite{fort2019deep, wilson2020bayesian}. This may allow the neural network to generalize better to complex acoustic scenarios.

To show the impact of modeling predictive uncertainty that combines both aleatoric and epistemic uncertainties on speech enhancement performance, we use the same set of models as described in Section~\ref{analysispredictiveuncertainty}. We take the average of estimates as in~\eqref{mean_wf_var_pre} and~\eqref{mean_amap} and obtain two speech estimates, called DE-Aleatoric-WF and DE-Aleatoric-AMAP respectively. They both provide better ESTOI and SI-SDR scores than the baselines, the epistemic uncertainty-only model, and the aleatoric uncertainty-only model, especially at low SNRs. Moreover, DE-Aleatoric-AMAP yields higher scores in \ac{PESQ} likely due to the uncertainty-dependent regularization and exploration of multiple modes in the function space. This indicates that combining the model that accounts for aleatoric uncertainty with the ensemble-based method can take advantage of the benefits of both approaches and further improve the performance. Overall, the evaluation results across different datasets show that quantifying uncertainty in neural network-based speech enhancement leads to a considerable improvement in enhancement performance over the baseline models.

\section{Conclusion}
\label{sec:conclusion}
\noindent In this paper, besides estimating clean speech, we quantified predictive uncertainty in neural network-based speech enhancement. For this, aleatoric uncertainty, which describes inherent uncertainty in data, and epistemic uncertainty, which accounts for uncertainty of the model, were captured and analyzed in a joint framework. We investigated the reliability of uncertainty estimates from different sources, and how it affects the enhancement performance. Our proposed hybrid loss function based on \ac{MAP} inference of complex spectral coefficients and an \ac{AMAP} estimator of spectral magnitudes has demonstrated the effectiveness in modeling aleatoric uncertainty. In addition, the proposed scheme provided a principled way to create a noise-removing mask that explicitly incorporates uncertainty to further improve speech enhancement performance. The evaluation results on different datasets have shown increased generalization capacities when modeling aleatoric uncertainty. 

To empirically approximate the predictive distribution and capture epistemic uncertainty, we employed two Bayesian deep learning methods, MC dropout and Deep ensembles. We showed that Deep ensembles not only provide more accurate estimates of epistemic uncertainty than MC dropout, but also lead to more prominent improvements in speech enhancement. A reason may be that Deep ensembles can potentially converge to different local minima in the loss landscape due to random initialization. Furthermore, we combined the proposed hybrid function with Deep ensembles to quantify overall prediction uncertainty, which reflects both data uncertainty and model uncertainty. An analysis using sparsification plots showed that combining different types of uncertainties further improves the reliability of predictive uncertainty estimation, indicating the complementary nature of the two sources of uncertainty. Finally, our experiments indicated that the performance of clean speech estimation can be considerably improved over the baselines while additionally obtaining predictive uncertainty estimates.

In summary, this work investigated capturing predictive uncertainty in neural network-based speech enhancement and showed the noticeable benefits of modeling uncertainty for clean speech estimation. Uncertainty can indicate the algorithm's confidence in the output in the absence of ground truth, which is essential for assessing the reliability of speech estimates. With this work, we hope to enlighten discussions on modeling uncertainty in the speech enhancement task, while facilitating future research on how to take advantage of uncertainty.

\balance

\bibliographystyle{IEEEtran}

\bibliography{refs}

\end{document}

%% file: block_diagram.tikz
    \tikzstyle{normalnode}=[dspsquare, inner sep=1mm]%
    \tikzstyle{bignode}=[normalnode, minimum height=2cm, align=center]
    \tikzstyle{smallnode}=[normalnode]
    \tikzstyle{legendnode}=[normalnode, minimum height=0.2cm, minimum width=0.5cm, inner sep=0]
    \tikzstyle{loss}=[normalnode, fill=blue!30, minimum width=1.3cm]
    \tikzstyle{connarrow}=[dspconn]
    \begin{tikzpicture}[node distance=1cm and 1.5cm]
        \node[smallnode] (Xmag) {$\left| \cdot \right|$};
        \node[left=of Xmag] (Xin) {};
        \node[bignode, fill=orange!30, right=of Xmag] (NN) {\footnotesize DNN};
        \node[bignode, fill=green!30, right=of NN] (AMAP) {\footnotesize AMAP};
        \node[dspmixer, right=of AMAP] (mul) {};
        \node[right=of mul] (Smag_hat) {};

        \node[dspnodefull, right=1cm of Xmag] (Xmag_split) {};

        \draw[connarrow] (Xin) -- node[above, align=center] {Noisy speech\\$X$} (Xmag);
        \draw[connarrow] (Xmag) -- node[above, xshift=-0.2cm] {$\left| X \right|$} (NN);
        \draw[connarrow] ([yshift=0.5cm]NN.east) -- node[above,xshift=-0.35cm] {$W^{\text{WF}}$} node[dspnodefull,xshift=0.2cm] (w_split) {} ([yshift=0.5cm]AMAP.west);
        \draw[connarrow] ([yshift=-0.5cm]NN.east) -- node[below,xshift=-0.5cm] {$\lambda$} node[dspnodefull,xshift=-0.2cm] (lambda_split) {} node[] (lambda_dummy) {} ([yshift=-0.5cm]AMAP.west);
        \draw[connarrow] (AMAP) -- node[above] {$W^{\text{AMAP}}$} (mul);
        \draw[connarrow] (mul) -- node[above] {$| \widehat{S} |$} node[dspnodefull] (Smag_hat_split) {} (Smag_hat);
        
        \draw[connarrow, rounded corners=1mm] (Xmag_split) -- +(0, 1.5) -| (mul);

        \node[loss, below=of lambda_dummy] (unc_loss) {$\mathcal{L}_{p(S|X
        )}$};
        \draw[connarrow, dotted] (w_split) -- ([xshift=0.2cm]unc_loss.north);
        \draw[connarrow, dotted] (lambda_split) -- ([xshift=-0.2cm]unc_loss.north);

        \node[loss] at (Smag_hat_split |- unc_loss) (sisdr_loss) {$\mathcal{L}_{\text{SI-SDR}}$};
        \node[legendnode, dotted, below=0.5cm of Smag_hat_split, inner sep=0.5mm] (istft) {iSTFT};
        \draw[connarrow, dotted] (Smag_hat_split) -- (istft.north);
        \draw[connarrow, dotted] (istft.south) -- (sisdr_loss.north);

        \node[legendnode, fill=orange!30, label=right:{\small Trainable}, xshift=2cm, yshift=1.3cm] at (sisdr_loss |- unc_loss) (legend_train) {};
        \node[legendnode, fill=green!30, label=right:{\small Non-trainable}, below=0.3cm of legend_train] (legend_fixed) {};.
        \node[legendnode, fill=blue!30, label=right:{\small Loss terms}, below=0.3cm of legend_fixed] (legend_loss) {};

    \end{tikzpicture}

%% file: main.bbl
\def\ICASSP{IEEE Int. Conf. Acoustics, Speech, Signal Proc.
  (ICASSP)}\def\INTERSPEECH{Interspeech}\def\IEEETASL{IEEE Trans. Audio,
  Speech, Language Proc.}\def\IEEETASLACM{IEEE/ACM Trans. Audio, Speech,
  Language Proc.}\def\TSAP{IEEE Trans. on Speech and Audio
  Proc.}\def\TASSP{IEEE Trans. on Acoustics, Speech and Signal
  Proc.}\def\TSP{IEEE Trans. on Signal Proc.}\def\MLSP{ Int. Workshop on
  Machine Learning for Signal Processing (MLSP)}\def\IWAENC{Int. Workshop on
  Acoustic Echo and Noise Control (IWAENC)}\def\EUSIPCO{Proc. Euro. Signal
  Proc. Conf. (EUSIPCO)}\def\WASPAA{IEEE Workshop Applications Signal Proc.
  Audio, Acoustics (WASPAA)}\def\ICLR{Int. Conf. Learning Repr.
  (ICLR)}\def\ICML{Int. Conf. Machine Learning (ICML)}\def\ICCV{IEEE/CVF Int.
  Conf. on Computer Vision (ICCV)}\def\CVPR{IEEE Conf. on Computer Vision and
  Pattern Recognition}\def\NIPS{Advances in Neural Information Proc. Systems}
\begin{thebibliography}{10}
\providecommand{\url}[1]{#1}
\csname url@samestyle\endcsname
\providecommand{\newblock}{\relax}
\providecommand{\bibinfo}[2]{#2}
\providecommand{\BIBentrySTDinterwordspacing}{\spaceskip=0pt\relax}
\providecommand{\BIBentryALTinterwordstretchfactor}{4}
\providecommand{\BIBentryALTinterwordspacing}{\spaceskip=\fontdimen2\font plus
\BIBentryALTinterwordstretchfactor\fontdimen3\font minus
  \fontdimen4\font\relax}
\providecommand{\BIBforeignlanguage}[2]{{%
\expandafter\ifx\csname l@#1\endcsname\relax
\typeout{** WARNING: IEEEtran.bst: No hyphenation pattern has been}%
\typeout{** loaded for the language `#1'. Using the pattern for}%
\typeout{** the default language instead.}%
\else
\language=\csname l@#1\endcsname
\fi
#2}}
\providecommand{\BIBdecl}{\relax}
\BIBdecl

\bibitem{timo2012transac}
T.~{Gerkmann} and R.~C. {Hendriks}, ``Unbiased {MMSE}-based noise power
  estimation with low complexity and low tracking delay,'' \emph{\IEEETASL},
  vol.~20, no.~4, pp. 1383--1393, 2012.

\bibitem{timowienerfiltering2018}
T.~Gerkmann and E.~Vincent, ``Spectral masking and filtering,'' in \emph{Audio
  Source Separation and Speech Enhancement}, E.~Vincent, T.~Virtanen, and
  S.~Gannot, Eds.\hskip 1em plus 0.5em minus 0.4em\relax Wiley, 2018, pp.
  65--85.

\bibitem{mmse1984}
Y.~Ephraim and D.~Malah, ``Speech enhancement using a minimum-mean square error
  short-time spectral amplitude estimator,'' \emph{\TASSP}, vol.~32, no.~6, pp.
  1109--1121, 1984.

\bibitem{wolfe2003efficient}
P.~J. Wolfe and S.~J. Godsill, ``Efficient alternatives to the ephraim and
  malah suppression rule for audio signal enhancement,'' \emph{EURASIP Journal
  on Advances in Signal Proc.}, vol. 2003, no.~10, pp. 1--9, 2003.

\bibitem{Ioannis2006supergaussian}
I.~Andrianakis and P.~White, ``{MMSE} speech spectral amplitude estimators with
  {Chi} and {Gamma} speech priors,'' in \emph{\ICASSP}, May 2006, pp. III--III.

\bibitem{Breithaupt2008pamaetrizedMMSE}
C.~Breithaupt, M.~Krawczyk, and R.~Martin, ``Parameterized {MMSE} spectral
  magnitude estimation for the enhancement of noisy speech,'' in
  \emph{\ICASSP}, Mar. 2008, pp. 4037--4040.

\bibitem{martin2001}
R.~Martin, ``Noise power spectral density estimation based on optimal smoothing
  and minimum statistics,'' \emph{\TSAP}, vol.~9, no.~5, pp. 504--512, 2001.

\bibitem{marvin2020spp}
M.~Tammen, D.~Fischer, B.~T. Meyer, and S.~Doclo, ``{DNN}-based speech presence
  probability estimation for multi-frame single-microphone speech
  enhancement,'' in \emph{\ICASSP}, May 2020, pp. 191--195.

\bibitem{mfmvdr2012huang}
Y.~A. Huang and J.~Benesty, ``A multi-frame approach to the frequency-domain
  single-channel noise reduction problem,'' \emph{\IEEETASL}, vol.~20, no.~4,
  pp. 1256--1269, 2012.

\bibitem{suhadi2011}
S.~Suhadi, C.~Last, and T.~Fingscheidt, ``A data-driven approach to a priori
  snr estimation,'' \emph{\IEEETASL}, vol.~19, no.~1, pp. 186--195, 2011.

\bibitem{deepmmse2020}
Q.~Zhang, A.~Nicolson, M.~Wang, K.~K. Paliwal, and C.~Wang, ``Deep{MMSE}: A
  deep learning approach to {MMSE}-based noise power spectral density
  estimation,'' \emph{\IEEETASLACM}, vol.~28, pp. 1404--1415, 2020.

\bibitem{bandovae}
Y.~{Bando}, M.~{Mimura}, K.~{Itoyama}, K.~{Yoshii}, and T.~{Kawahara},
  ``Statistical speech enhancement based on probabilistic integration of
  variational autoencoder and non-negative matrix factorization,'' in
  \emph{\ICASSP}, Apr. 2018, pp. 716--720.

\bibitem{Takuya2021}
T.~Fujimura, Y.~Koizumi, K.~Yatabe, and R.~Miyazaki, ``Noisy-target training: A
  training strategy for dnn-based speech enhancement without clean speech,'' in
  \emph{European Signal Proc. Conf. (EUSIPCO)}, 2021, pp. 436--440.

\bibitem{simonvae}
S.~{Leglaive}, L.~{Girin}, and R.~{Horaud}, ``A variance modeling framework
  based on variational autoencoders for speech enhancement,'' in \emph{\MLSP},
  Sep. 2018, pp. 1--6.

\bibitem{carbajal2021classifier}
G.~Carbajal, J.~Richter, and T.~Gerkmann, ``Guided variational autoencoder for
  speech enhancement with a supervised classifier,'' in \emph{\ICASSP}, Jun.
  2021, pp. 681--685.

\bibitem{fang2021variational}
H.~Fang, G.~Carbajal, S.~Wermter, and T.~Gerkmann, ``Variational autoencoder
  for speech enhancement with a noise-aware encoder,'' in \emph{\ICASSP}, Jun.
  2021, pp. 676--680.

\bibitem{carbajal2021disentangle}
G.~Carbajal, J.~Richter, and T.~Gerkmann, ``Disentanglement learning for
  variational autoencoders applied to audio-visual speech enhancement,'' in
  \emph{\WASPAA}, Oct. 2021, pp. 126--130.

\bibitem{wang2018supervised}
D.~Wang and J.~Chen, ``Supervised speech separation based on deep learning: An
  overview,'' \emph{\IEEETASLACM}, vol.~26, no.~10, pp. 1702--1726, 2018.

\bibitem{braun2021loss}
S.~Braun and I.~Tashev, ``A consolidated view of loss functions for supervised
  deep learning-based speech enhancement,'' in \emph{Int. Conf. on
  Telecommunications and Signal Proc. (TSP)}, Jul. 2021, pp. 72--76.

\bibitem{richter2020speech}
J.~Richter, G.~Carbajal, and T.~Gerkmann, ``Speech enhancement with stochastic
  temporal convolutional networks,'' in \emph{\INTERSPEECH}, Oct. 2020, pp.
  4516--4520.

\bibitem{leglaive2020recurrent}
S.~Leglaive, X.~Alameda-Pineda, L.~Girin, and R.~Horaud, ``A recurrent
  variational autoencoder for speech enhancement,'' in \emph{\ICASSP}, May
  2020, pp. 371--375.

\bibitem{bie2022unsupervised}
X.~Bie, S.~Leglaive, X.~Alameda-Pineda, and L.~Girin, ``Unsupervised speech
  enhancement using dynamical variational autoencoders,'' \emph{\IEEETASLACM},
  vol.~30, pp. 2993--3007, 2022.

\bibitem{sadeghi2020audio}
M.~Sadeghi, S.~Leglaive, X.~Alameda-Pineda, L.~Girin, and R.~Horaud,
  ``Audio-visual speech enhancement using conditional variational
  auto-encoders,'' \emph{\IEEETASLACM}, vol.~28, pp. 1788--1800, 2020.

\bibitem{fu2021metricgan+}
S.-W. Fu, C.~Yu, T.-A. Hsieh, P.~Plantinga, M.~Ravanelli, X.~Lu, and Y.~Tsao,
  ``{MetricGAN+: An improved version of MetricGAN for speech enhancement},'' in
  \emph{\INTERSPEECH}, Aug. 2021, pp. 201--205.

\bibitem{fu2022metricgan}
S.-W. Fu, C.~Yu, K.-H. Hung, M.~Ravanelli, and Y.~Tsao, ``{MetricGAN-U:
  Unsupervised speech enhancement/dereverberation based only on
  noisy/reverberated speech},'' in \emph{\ICASSP}, May 2022, pp. 7412--7416.

\bibitem{lu2022conditional}
Y.-J. Lu, Z.-Q. Wang, S.~Watanabe, A.~Richard, C.~Yu, and Y.~Tsao,
  ``Conditional diffusion probabilistic model for speech enhancement,'' in
  \emph{\ICASSP}, May 2022, pp. 7402--7406.

\bibitem{welker2022speech}
S.~Welker, J.~Richter, and T.~Gerkmann, ``Speech enhancement with score-based
  generative models in the complex {STFT} domain,'' in \emph{\INTERSPEECH},
  Sep. 2022, pp. 2928--2932.

\bibitem{rehr2021snr}
R.~Rehr and T.~Gerkmann, ``{SNR}-based features and diverse training data for
  robust {DNN}-based speech enhancement,'' \emph{\IEEETASLACM}, vol.~29, pp.
  1937--1949, 2021.

\bibitem{nat2016}
A.~Kumar and D.~Florencio, ``Speech enhancement in multiple-noise conditions
  using deep neural networks,'' in \emph{\INTERSPEECH}, Sep. 2016, p.
  3738–3742.

\bibitem{largescale2016}
J.~Chen, Y.~Wang, S.~E. Yoho, D.~Wang, and E.~W. Healy, ``Large-scale training
  to increase speech intelligibility for hearing-impaired listeners in novel
  noises,'' \emph{The Journal of the Acoustical Society of America}, vol. 139,
  no.~5, pp. 2604--2612, 2016.

\bibitem{aleaepicML2021}
E.~H{\"u}llermeier and W.~Waegeman, ``Aleatoric and epistemic uncertainty in
  machine learning: An introduction to concepts and methods,'' \emph{Machine
  Learning}, vol. 110, no.~3, pp. 457--506, 2021.

\bibitem{aleaepic2009}
A.~Der~Kiureghian and O.~Ditlevsen, ``{Aleatory or epistemic? Does it
  matter?}'' \emph{Structural Safety}, vol.~31, no.~2, pp. 105--112, 2009.

\bibitem{kendall2017uncertainties}
A.~Kendall and Y.~Gal, ``What uncertainties do we need in {Bayesian} deep
  learning for computer vision?'' \emph{\NIPS}, vol.~30, 2017.

\bibitem{depeweg2018decomposition}
S.~Depeweg, J.-M. Hernandez-Lobato, F.~Doshi-Velez, and S.~Udluft,
  ``Decomposition of uncertainty in {Bayesian} deep learning for efficient and
  risk-sensitive learning,'' in \emph{\ICML}, Jul. 2018, pp. 1184--1193.

\bibitem{pearce2018high}
T.~Pearce, A.~Brintrup, M.~Zaki, and A.~Neely, ``High-quality prediction
  intervals for deep learning: A distribution-free, ensembled approach,'' in
  \emph{\ICML}, Jul. 2018, pp. 4075--4084.

\bibitem{lakshminarayanan2017simple}
B.~Lakshminarayanan, A.~Pritzel, and C.~Blundell, ``Simple and scalable
  predictive uncertainty estimation using deep ensembles,'' \emph{\NIPS},
  vol.~30, 2017.

\bibitem{gustafsson2020evaluating}
F.~K. Gustafsson, M.~Danelljan, and T.~B. Schon, ``Evaluating scalable
  {Bayesian} deep learning methods for robust computer vision,'' in \emph{\CVPR
  \, Workshops}, Jun. 2020, pp. 318--319.

\bibitem{ilg2018uncertainty}
E.~Ilg, O.~Cicek, S.~Galesso, A.~Klein, O.~Makansi, F.~Hutter, and T.~Brox,
  ``Uncertainty estimates and multi-hypotheses networks for optical flow,'' in
  \emph{European Conf. on Computer Vision (ECCV)}, Sep. 2018, pp. 652--667.

\bibitem{chai2019generalizedgaussian}
L.~Chai, J.~Du, Q.-F. Liu, and C.-H. Lee, ``Using generalized {Gaussian}
  distributions to improve regression error modeling for deep learning-based
  speech enhancement,'' \emph{\IEEETASLACM}, vol.~27, no.~12, pp. 1919--1931,
  2019.

\bibitem{kinishita2017mdn}
K.~Kinoshita, M.~Delcroix, A.~Ogawa, T.~Higuchi, and T.~Nakatani, ``Deep
  mixture density network for statistical model-based feature enhancement,'' in
  \emph{\ICASSP}, Mar. 2017, pp. 251--255.

\bibitem{Siniscalchi2021distributionalloss}
S.~M. Siniscalchi, ``Vector-to-vector regression via distributional loss for
  speech enhancement,'' \emph{IEEE Signal Processing Letters}, vol.~28, pp.
  254--258, 2021.

\bibitem{gal2016dropout}
Y.~Gal and Z.~Ghahramani, ``Dropout as a {Bayesian} approximation: Representing
  model uncertainty in deep learning,'' in \emph{\ICML}, Jun. 2016, pp.
  1050--1059.

\bibitem{welling2011bayesian}
M.~Welling and Y.~W. Teh, ``Bayesian learning via stochastic gradient
  {Langevin} dynamics,'' in \emph{\ICML}, Jun. 2011, pp. 681--688.

\bibitem{chen2014stochastic}
T.~Chen, E.~Fox, and C.~Guestrin, ``Stochastic gradient {Hamiltonian Monte
  Carlo},'' in \emph{\ICML}, Jun. 2014, pp. 1683--1691.

\bibitem{blundell2015weight}
C.~Blundell, J.~Cornebise, K.~Kavukcuoglu, and D.~Wierstra, ``Weight
  uncertainty in neural network,'' in \emph{\ICML}, Jun. 2015, pp. 1613--1622.

\bibitem{gal2015bayesian}
Y.~Gal and Z.~Ghahramani, ``Bayesian convolutional neural networks with
  {B}ernoulli approximate variational inference,'' in \emph{\ICLR \ workshop
  track}, 2016.

\bibitem{nixon2020bootstrapped}
J.~Nixon, B.~Lakshminarayanan, and D.~Tran, ``Why are bootstrapped deep
  ensembles not better?'' in \emph{``I Can't Believe It's Not Better!'' Neural
  Information Proc. Systems workshop}, 2020.

\bibitem{fort2019deep}
S.~Fort, H.~Hu, and B.~Lakshminarayanan, ``Deep ensembles: A loss landscape
  perspective,'' in \emph{``Bayesian Deep Learning'' Neural Information Proc.
  Systems workshop}, 2019.

\bibitem{wilson2020bayesian}
A.~G. Wilson and P.~Izmailov, ``Bayesian deep learning and a probabilistic
  perspective of generalization,'' \emph{\NIPS}, vol.~33, pp. 4697--4708, 2020.

\bibitem{dropout2014}
N.~Srivastava, G.~Hinton, A.~Krizhevsky, I.~Sutskever, and R.~Salakhutdinov,
  ``Dropout: {A} simple way to prevent neural networks from overfitting,''
  \emph{The Journal of Machine Learning Research}, vol.~15, no.~1, pp.
  1929--1958, 2014.

\bibitem{mcdropoutser2020}
K.~Sridhar and C.~Busso, ``Modeling uncertainty in predicting emotional
  attributes from spontaneous speech,'' in \emph{\ICASSP}, May 2020, pp.
  8384--8388.

\bibitem{navin2022}
N.~Raj~Prabhu, G.~Carbajal, N.~Lehmann-Willenbrock, and G.~Timo, ``End-to-end
  label uncertainty modeling for speech-based arousal recognition using
  {Bayesian} neural networks,'' in \emph{\INTERSPEECH}, Sep. 2022, pp.
  151--155.

\bibitem{bbpsr2019}
S.~Braun and S.-C. Liu, ``Parameter uncertainty for end-to-end speech
  recognition,'' in \emph{\ICASSP}, May 2019, pp. 5636--5640.

\bibitem{mcdropoutsr202}
S.~Khurana, N.~Moritz, T.~Hori, and J.~L. Roux, ``Unsupervised domain
  adaptation for speech recognition via uncertainty driven self-training,'' in
  \emph{\ICASSP}, Jun. 2021, pp. 6553--6557.

\bibitem{asrdropout2019}
A.~Vyas, P.~Dighe, S.~Tong, and H.~Bourlard, ``Analyzing uncertainties in
  speech recognition using dropout,'' in \emph{\ICASSP}, May 2019, pp.
  6730--6734.

\bibitem{kendall2017bsegnet}
K.~Alex, B.~Vijay, and C.~Roberto, ``Bayesian {SegNet}: Model uncertainty in
  deep convolutional encoder-decoder architectures for scene understanding,''
  in \emph{Proc. of the British Machine Vision Conf. (BMVC)}, Sep 2017, pp.
  57.1--57.12.

\bibitem{ensemblejohnatan2013}
J.~Le~Roux, S.~Watanabe, and J.~R. Hershey, ``Ensemble learning for speech
  enhancement,'' in \emph{\WASPAA}, Oct. 2013, pp. 1--4.

\bibitem{zhang2017multi}
H.~Zhang, X.~Zhang, and G.~Gao, ``Multi-target ensemble learning for monaural
  speech separation,'' in \emph{\INTERSPEECH}, Aug. 2017, pp. 1958--1962.

\bibitem{fang2022uncertainty}
H.~Fang, T.~Peer, S.~Wermter, and T.~Gerkmann, ``Integrating statistical
  uncertainty into neural network-based speech enhancement,'' in
  \emph{\ICASSP}, May 2022, pp. 386--390.

\bibitem{fevotte2011algorithms}
C.~F{\'e}votte and J.~Idier, ``Algorithms for nonnegative matrix factorization
  with the $\beta$-divergence,'' \emph{Neural computation}, vol.~23, no.~9, pp.
  2421--2456, 2011.

\bibitem{pitfallnll2022}
M.~Seitzer, A.~Tavakoli, D.~Antic, and G.~Martius, ``On the pitfalls of
  heteroscedastic uncertainty estimation with probabilistic neural networks,''
  in \emph{\ICLR}, Apr. 2022.

\bibitem{varvariance2020}
A.~Stirn and D.~A. Knowles, ``Variational variance: Simple, reliable,
  calibrated heteroscedastic noise variance parameterization,''
  \emph{arXiv:2006.04910}, 2020.

\bibitem{reliableNLL2019}
N.~Skafte, M.~J{\o}rgensen, and S.~Hauberg, ``Reliable training and estimation
  of variance networks,'' \emph{\NIPS}, vol.~32, 2019.

\bibitem{macaulayMalpass1980}
R.~McAulay and M.~Malpass, ``Speech enhancement using a soft-decision noise
  suppression filter,'' \emph{\TASSP}, vol.~28, no.~2, pp. 137--145, 1980.

\bibitem{le2019sdr}
J.~Le~Roux, S.~Wisdom, H.~Erdogan, and J.~R. Hershey, ``{SDR}--half-baked or
  well done?'' in \emph{\ICASSP}, May 2019, pp. 626--630.

\bibitem{heitkaemper2020demystifying}
J.~Heitkaemper, D.~Jakobeit, C.~Boeddeker, L.~Drude, and R.~Haeb-Umbach,
  ``{Demystifying TasNet: A dissecting approach},'' in \emph{\ICASSP}, May
  2020, pp. 6359--6363.

\bibitem{Neal1995BayesianNN}
R.~M. Neal, ``Bayesian learning for neural networks,'' {PhD} thesis, University
  of Toronto, 1995.

\bibitem{reddy2020interspeech}
C.~K. Reddy, V.~Gopal, R.~Cutler, E.~Beyrami, R.~Cheng, H.~Dubey,
  S.~Matusevych, R.~Aichner, A.~Aazami, S.~Braun, P.~Rana, S.~Srinivasan, and
  J.~Gehrke, ``The {{INTERSPEECH}} 2020 {{Deep Noise Suppression Challenge}}:
  {{Datasets}}, {{Subjective Testing Framework}}, and {{Challenge Results}},''
  in \emph{\INTERSPEECH}, May 2020, pp. 2492--2496.

\bibitem{pirker2011pitch}
G.~Pirker, M.~Wohlmayr, S.~Petrik, and F.~Pernkopf, ``A pitch tracking corpus
  with evaluation on multipitch tracking scenario,'' in \emph{\INTERSPEECH},
  Aug. 2011, pp. 1509--1512.

\bibitem{garofolo1993csr}
J.~Garofolo, D.~Graff, D.~Paul, and D.~Pallett, ``{CSR-I} ({WSJ0}) {S}ennheiser
  {LDC93S6B},'' \emph{Web Download. Philadelphia: Linguistic Data Consortium},
  1993.

\bibitem{chime3dataset}
J.~Barker, R.~Marxer, E.~Vincent, and S.~Watanabe, ``The third ‘{CHiME}’
  speech separation and recognition challenge: Dataset, task and baselines,''
  in \emph{2015 IEEE Workshop on Automatic Speech Recognition and Understanding
  (ASRU)}, Dec. 2015, pp. 504--511.

\bibitem{Jansson2017SingingVS}
A.~Jansson, E.~J. Humphrey, N.~Montecchio, R.~M. Bittner, A.~Kumar, and
  T.~Weyde, ``Singing voice separation with deep {U-Net} convolutional
  networks,'' in \emph{ISMIR}, Oct. 2017.

\bibitem{tan2018convolutional}
K.~Tan and D.~Wang, ``A convolutional recurrent neural network for real-time
  speech enhancement,'' in \emph{\INTERSPEECH}, Sep. 2018, pp. 3229--3233.

\bibitem{ulyanov2017improved}
D.~Ulyanov, A.~Vedaldi, and V.~Lempitsky, ``Improved texture networks:
  Maximizing quality and diversity in feed-forward stylization and texture
  synthesis,'' in \emph{\CVPR}, Jul. 2017, pp. 6924--6932.

\bibitem{adamkinma}
D.~P. Kingma and J.~Ba, ``Adam: A method for stochastic optimization,''
  \emph{\ICLR}, Dec. 2014.

\bibitem{wannenwetsch2017probflow}
A.~S. Wannenwetsch, M.~Keuper, and S.~Roth, ``Probflow: Joint optical flow and
  uncertainty estimation,'' in \emph{\ICCV}, Oct. 2017, pp. 1173--1182.

\bibitem{pesq}
A.~Rix, J.~Beerends, M.~Hollier, and A.~Hekstra, ``Perceptual evaluation of
  speech quality ({PESQ})-a new method for speech quality assessment of
  telephone networks and codecs,'' in \emph{\ICASSP}, May 2001, pp. 749--752.

\bibitem{estoi}
J.~{Jensen} and C.~H. {Taal}, ``An algorithm for predicting the intelligibility
  of speech masked by modulated noise maskers,'' \emph{\IEEETASLACM}, vol.~24,
  no.~11, pp. 2009--2022, 2016.

\end{thebibliography}
